\documentclass[aps,prxquantum,reprint,twocolumn,superscriptaddress,floatfix,nofootinbib,longbibliography]{revtex4-2}
\usepackage{epsfig,amsmath,amssymb,color,dsfont,upgreek,physics}
\usepackage{mathrsfs}
\usepackage{mathtools}
\usepackage{bbold}
\usepackage{float}
\usepackage[bookmarks=true,colorlinks,linkcolor=OrangeRed,urlcolor=NavyBlue,citecolor=RoyalBlue]{hyperref}
\usepackage[dvipsnames]{xcolor}
\usepackage{cleveref}
\usepackage{orcidlink}

\definecolor{mygold}{rgb}{0.5,0.6,0.7}

\definecolor{mypurple}{rgb}{0.49,0.18,0.56}

\definecolor{mygreen}{rgb}{0,0.5,0}
\definecolor{myred}{rgb}{0.7,0,0}
\definecolor{myblue}{rgb}{0,0,1}

\usepackage{changes}

\usepackage{tikz}
 \usetikzlibrary{calc}
\usetikzlibrary{arrows.meta}
\usetikzlibrary{decorations.pathreplacing}
\usepackage{graphicx}
\usepackage{comment}
\usetikzlibrary{decorations.markings}
\tikzset{two arrows/.style={
    postaction={decorate},
  decoration={
    markings,
    mark=at position 0.45 with {\arrow[scale=1.6]{Latex[length=5mm, width=4mm]}},
    mark=at position 0.94 with {\arrow[scale=1.6]{Latex[length=5mm, width=4mm]}}
  }
}}
\tikzset{
  mid arrow/.style={
    postaction={decorate},
    decoration={
      markings,
      mark=at position 0.7 with {\arrow[scale=1.6]{Latex[length=5mm, width=4mm]}}
    }
  }
}
\tikzset{
  out arrow/.style={
    postaction={decorate},
    decoration={
      markings,
      mark=at position 0.85 with {\arrow[scale=1.6]{Latex[length=5mm, width=4mm]}}
    }
  }
}

\newcommand{\well}[2]{%
  \begin{tikzpicture}[scale=0.4, baseline={(0,-0.2)}]
    \draw[thick,domain=-1.2:1.7,smooth,samples=200]
      plot(\x,{0.9*((\x)^2 - 1)^2 - 0.5*\x^3});

    \ifnum#1>0
      \foreach \i in {1,...,#1}
        \filldraw[red] (-0.75, 0.5*\i + 0.4) circle (3pt);
    \fi

    \ifnum#2>0
      \foreach \i in {1,...,#2}
        \filldraw[red] (1.1, 0.5*\i - 0.2) circle (3pt);
    \fi
  \end{tikzpicture}%
}
\NewDocumentCommand{\Op}{m o o s}{
  \hat{#1}
  \IfValueT{#2}{_{#2}}
  \IfValueTF{#3}{
    ^{#3\IfBooleanT{#4}{\dagger}}
  }{
    \IfBooleanT{#4}{^{\dagger}}
  }
}

\NewDocumentCommand{\state}{oo}{
  \ket{
    \psi
    \IfValueT{#1}{_{#1}}
    \IfValueT{#2}{^{#2}}
  }
}
\let\groupStyle\mathrm
\newcommand{\UN}{\groupStyle{U}}
\newcommand{\SUN}{\groupStyle{SU}}
\def\qlm{\mathrm{QLM}}
\def\bhm{\mathrm{BHM}}

\begin{document}
\title{Towards $2+1$D quantum electrodynamics on a cold-atom quantum simulator}

\author{Peter Majcen${}^{\orcidlink{0009-0009-7020-7246}}$}
\thanks{These authors contributed equally to this work.}
\affiliation{Dipartimento di Fisica e Astronomia ``G. Galilei'', Universit\`a di Padova, I-35131 Padova, Italy}
\affiliation{Padua Quantum Technologies Research Center, Universit\`a degli Studi di Padova}
\affiliation{Istituto Nazionale di Fisica Nucleare (INFN), Sezione di Padova, I-35131 Padova, Italy}

\author{Jesse J.~Osborne${}^{\orcidlink{0000-0003-0415-0690}}$}
\thanks{These authors contributed equally to this work.}
\affiliation{Max Planck Institute of Quantum Optics, 85748 Garching, Germany}
\affiliation{Department of Physics and Arnold Sommerfeld Center for Theoretical Physics (ASC),\\ Ludwig Maximilian University of Munich, 80333 Munich, Germany}
\affiliation{Munich Center for Quantum Science and Technology (MCQST), 80799 Munich, Germany}

\author{Philipp Hauke${}^{\orcidlink{0000-0002-0414-1754}}$}
\affiliation{Pitaevskii BEC Center and Department of Physics, University of Trento, I-38123 Trento, Italy}
\affiliation{INFN-TIFPA, Trento Institute for Fundamental Physics and Applications, Trento, Italy}

\author{Bing Yang${}^{\orcidlink{0000-0002-8379-9289}}$}
\affiliation{Department of Physics, Southern University of Science and Technology, Shenzhen 518055, China}

\author{Simone Montangero${}^{\orcidlink{0000-0002-8882-2169}}$}
\affiliation{Dipartimento di Fisica e Astronomia ``G. Galilei'', Universit\`a di Padova, I-35131 Padova, Italy}
\affiliation{Padua Quantum Technologies Research Center, Universit\`a degli Studi di Padova}
\affiliation{Istituto Nazionale di Fisica Nucleare (INFN), Sezione di Padova, I-35131 Padova, Italy}

\author{Jad C.~Halimeh${}^{\orcidlink{0000-0002-0659-7990}}$}
\email{jad.halimeh@lmu.de}
\affiliation{Department of Physics and Arnold Sommerfeld Center for Theoretical Physics (ASC),\\ Ludwig Maximilian University of Munich, 80333 Munich, Germany}
\affiliation{Max Planck Institute of Quantum Optics, 85748 Garching, Germany}
\affiliation{Munich Center for Quantum Science and Technology (MCQST), 80799 Munich, Germany}
\affiliation{Department of Physics, College of Science, Kyung Hee University, Seoul 02447, Republic of Korea}

\begin{abstract}
Cold atoms have become a powerful platform for quantum-simulating lattice gauge theories in higher spatial dimensions. 
However, such realizations have been restricted to the lowest possible truncations of the gauge field, which limit the connections one can make to lattice quantum electrodynamics. Here, we propose a feasible cold-atom quantum simulator of a $(2+1)$-dimensional $\UN(1)$ lattice gauge theory in a spin $S=1$ truncation, featuring dynamical matter and gauge fields.
We derive a mapping of this theory onto a bosonic computational basis, stabilized by an emergent gauge-protection mechanism through quantum Zeno dynamics.
The implementation is based on a single-species Bose--Hubbard model realized in a tilted optical superlattice.
This approach requires only moderate experimental resources already available in current ultracold-atom platforms.
Using infinite matrix product state simulations, we benchmark real-time dynamics under global quenches.
The results demonstrate faithful evolution of the target gauge theory and robust preservation of the gauge constraints.
Our work significantly advances the experimental prospects for simulating higher-dimensional lattice gauge theories using larger gauge-field truncations.
\end{abstract}

\date{\today} 
\maketitle
{\hypersetup{linkcolor=mygold}
\tableofcontents}

\section{Introduction}

Gauge theories play a central role in describing a wide range of physical phenomena, from low-energy phases of matter to the fundamental interactions of particle physics. 
At low energies, they provide the theoretical framework for emergent phenomena such as quantum spin liquids \cite{Savary2016QuantumSpinLiquids,Balents2010SpinLiquidsFrustrated,wen2004quantum} and the fractional quantum Hall effect \cite{Kleinert1989GaugeFieldsCondensed,Fradkin2013FieldTheoriesCondensed}. 
At high energies, gauge theories form the foundation of the Standard Model \cite{Weinberg1995QuantumTheoryFields,Gattringer2009QuantumChromodynamicsLattice,Cheng1984GaugeTheoryElementary}, including, for example, the Abelian $\UN(1)$ gauge theory of quantum electrodynamics (QED) with dynamical matter, and the non-Abelian $\SUN(3)$ gauge theory that describes the strong interaction.

The defining property of gauge theories is the principle of gauge invariance, which is a local symmetry and manifests itself as local constraints between matter and gauge degrees of freedom.
Although perturbative methods accurately describe many aspects of gauge theories, non-perturbative phenomena in interacting gauge theories in $3+1$D remain poorly understood.
By discretizing continuum gauge theories on a space-time (or spatial) lattice, one obtains lattice gauge theories (LGTs) \cite{Kogut1975HamiltonianFormulationWilsons,Kogut1979AnIntroductionToLatticeGaugeTheory,Rothe2012LatticeGaugeTheories}, which provide a framework to investigate such non-perturbative physics \cite{Wilson1974ConfinementQuarks,Wilson1977QuarksStringsLattice}.
Non-perturbative phenomena in lattice gauge theories are typically studied numerically using Monte Carlo techniques in the Lagrangian formulation \cite{Creutz1979MonteCarloStudy,Creutz1980MonteCarloStudy,Creutz1983MonteCarloComputations,Creutz1988LatticeGaugeTheory,Creutz1989LatticeGaugeTheories,montvay1994quantum,Kieu1994MonteCarloSimulations,Hackett2019DigitizingGaugeFields}. 
However, this approach encounters severe limitations when the Euclidean path-integral weight becomes complex, leading to the sign problem, an NP-hard challenge that hinders simulations at finite baryon density, as well as real-time dynamics \cite{Troyer2005ComputationalComplexityFundamental,Nagata2022FinitedensityLatticeQCD}.
However, one can also consider the Hamiltonian formulation of LGTs, which provides a quantum many-body framework \cite{Banuls2020SimulatingLatticeGauge, Dalmonte2016LatticeGaugeTheory, Zohar2015QuantumSimulationsLattice, Aidelsburger:2021mia,Zohar2021QuantumSimulationLattice, Bauer2023QuantumSimulationHighEnergy, Funcke2023ReviewQuantumComputing} and has led to intriguing insights into non-ergodic quantum many-body dynamics as a result of gauge invariance \cite{Smith2017DisorderFreeLocalization,Brenes2018ManyBodyLocalization,Smith2017AbsenceOfErgodicity,Karpov2021DisorderFreeLocalization,Sous2021PhononInducedDisorder,Chakraborty2022DisorderFreeLocalization,Halimeh2022EnhancingDisorderFreeLocalization,Surace2020LatticeGaugeTheories,Lang2022DisorderFreeLocalization,Desaules2023WeakErgodicityBreaking,Desaules2023ProminentQuantumManyBodyScars,Aramthottil2022ScarStates,Tarabunga2023ManyBodyMagic,Desaules2024ergodicitybreaking,Desaules2024MassAssistedLocalDeconfinement,Hudomal2022DrivingQuantumManyBodyScars,Jeyaretnam2025HilbertSpaceFragmentation,Smith2025Nonstabilizerness,Falcao2025nonstabilizerness,Esposito2025magicdiscretelatticegaugetheories,Ciavarella2025GenericHilbertSpaceFragmentation,Ciavarella:2025tdl,Steinegger2025GeometricFragmentationAnomalousThermalization,Ebner2024EntanglementEntropy,Halimeh2023robustquantummany,Iadecola2020QuantumManyBodyScar,Banerjee2021QuantumScarsZeroModes,Biswas2022ScarsFromProtectedZeroModes,Daniel2023BridgingQuantumCriticality,Sau2024sublatticescarsbeyond,Osborne2024QuantumManyBodyScarring,Budde2024QuantumManyBodyScars,Calajo2025QuantumManyBodyScarringNonAbelian,Hartse2025StabilizerScars}. This Hamiltonian approach complements Euclidean lattice approaches and enables direct access to real-time evolution. While this formulation of LGTs is commonly studied using classical methods, in particular those based on tensor networks \cite{Banuls2020ReviewNovelMethods,Zohar2015QuantumSimulationsLattice,Wiese2013UltracoldQuantumGases,Rigobello2021EntanglementGeneration$1+1mathrmD$,Schollwock2011DensitymatrixRenormalizationGroup,Orus2014PracticalIntroductionTensorNetworks,Montangero2018IntroductionTensorNetwork,Orus2019TensorNetworksComplex,Paeckel2019TimeevolutionMethodsMatrixproduct}, these are limited computationally for systems beyond one spatial dimension, large truncation levels of the gauge field, or for long-time evolution far from equilibrium due to rapid growth of entanglement entropy, despite recent progress in extending tensor-network approaches to higher spatial dimensions \cite{Cataldi2024Simulating2+1DSU2,Magnifico2021LatticeQuantumElectrodynamics}.
In this context, quantum simulators \cite{Bloch2008ManyBodyPhysics,Bloch2012QuantumSimulationUltracoldQuantumGases,Gross2017QuantumSimulations,Georgescu2014QuantumSimulation,Altman2021QuantumSimulators,Alexandrou:2025vaj} offer a promising route to overcoming classical computational barriers and are now approaching the limits of classical numerical methods \cite{Daley2022PracticalQuantumAdvantage,Scholl2021QuantumSimulation2D,Manovitz2025QuantumCoarseningCollective}. Currently, there is a huge drive to quantum-simulate high-energy physics phenomena \cite{Byrnes2006SimulatingLatticeGauge, Dalmonte2016LatticeGaugeTheory, Zohar2015QuantumSimulationsLattice, Aidelsburger:2021mia, Zohar2021QuantumSimulationLattice, 
Barata2022MediumInducedJetBroadening,Klco2022StandardModelPhysics,Barata2023QuantumSimulationInMediumQCDJets,Barata2023RealTimeDynamicsofHyperonSpin, Bauer2023QuantumSimulationHighEnergy, Bauer2023QuantumSimulationFundamental,
DiMeglio2024QuantumComputingHighEnergy, Cheng2024EmergentGaugeTheory, Halimeh2022StabilizingGaugeTheories, Cohen2021QuantumAlgorithmsTransport,Barata2025ProbingCelestialEnergy, Lee2025QuantumComputingEnergy, Turro2024ClassicalQuantumComputing,Halimeh2023ColdatomQuantumSimulators,Bauer2025EfficientUseQuantum,Halimeh2025QuantumSimulationOutofequilibrium}, with the holy grail being the \textit{ab initio} study of real-time dynamics of quantum chromodynamics (QCD). This would render such quantum simulators complementary venues to particle colliders. LGTs are at the forefront of this effort, with recent years producing a large collection of impressive experiments on analog and digital quantum hardware \cite{Martinez2016RealtimeDynamicsLattice, Klco2018QuantumclassicalComputationSchwinger,Gorg2019RealizationDensitydependentPeierls, Schweizer2019FloquetApproachZ2, Mil2020ScalableRealizationLocal, Yang2020ObservationGaugeInvariance, Wang2022ObservationEmergent$mathbbZ_2$, Su2023ObservationManybodyScarring, Zhou2022ThermalizationDynamicsGauge, Wang2023InterrelatedThermalizationQuantum, Zhang2025ObservationMicroscopicConfinement, Zhu2024ProbingFalseVacuum, Ciavarella2021TrailheadQuantumSimulation, Ciavarella2022PreparationSU3Lattice, Ciavarella2023QuantumSimulationLattice-1, Ciavarella2024QuantumSimulationSU3, 
Gustafson2024PrimitiveQuantumGates, Gustafson2024PrimitiveQuantumGates-1, Lamm2024BlockEncodingsDiscrete, Farrell2023PreparationsQuantumSimulations-1, Farrell2023PreparationsQuantumSimulations, 
Farrell2024ScalableCircuitsPreparing,
Farrell2024QuantumSimulationsHadron, Li2024SequencyHierarchyTruncation, Zemlevskiy2025ScalableQuantumSimulations, Lewis2019QubitModelU1, Atas2021SU2HadronsQuantum, ARahman:2022tkr, Atas2023SimulatingOnedimensionalQuantum, Mendicelli2023RealTimeEvolution, Kavaki2024SquarePlaquettesTriamond, Than2024PhaseDiagramQuantum, Angelides:2023noe, Gyawali2025ObservationDisorderfreeLocalization,  
Mildenberger2025Confinement$$mathbbZ_2$$Lattice, Schuhmacher2025ObservationHadronScattering, Davoudi2025QuantumComputationHadron, Saner2025RealTimeObservationAharonovBohm, Xiang2025RealtimeScatteringFreezeout, Wang2025ObservationInelasticMeson,li2025frameworkquantumsimulationsenergyloss,mark2025observationballisticplasmamemory,froland2025simulatingfullygaugefixedsu2,Hudomal2025ErgodicityBreakingMeetsCriticality,hayata2026onsetthermalizationqdeformedsu2,Cochran2025VisualizingDynamicsCharges, Gonzalez-Cuadra2025ObservationStringBreaking, Crippa2024AnalysisConfinementString, De2024ObservationStringbreakingDynamics, Liu2024StringBreakingMechanism, Alexandrou:2025vaj,Cobos2025RealTimeDynamics2+1D}. Despite this remarkable effort, these experimental realizations have been mostly restricted to the lowest two-level truncation of the gauge field, which severely limits extrapolation to the quantum field theory limit. We elucidate this point in the context of QED in the following.

QED plays a central role among gauge theories, both for theoretical understanding and for its experimental relevance.
To facilitate numerical and experimental studies, the quantum link formulation \cite{Chandrasekharan1997QuantumLinkModels,Wiese2013UltracoldQuantumGases,Kasper2017ImplementingQuantumElectrodynamics} of QED is employed, where gauge- and electric-field operators are encoded in spin-$S$ operators.
The resulting truncated version of QED recovers the full theory in the limit $S\rightarrow \infty$. At $S=1/2$, although certain qualitative features such as Coleman's phase transition are recovered \cite{Halimeh2022TuningtheTopologicaltheta}, the truncation is insufficient to capture the QED limit. However, at relatively small values of $S$, this limit can be faithfully approached at least in one spatial dimension \cite{Buyens2017FiniteRepresentationLatticeGaugeTheories,Zache2022TowardContinuumLimit,Halimeh2022achievingquantum}.
In Ref.~\cite{Osborne2025Scale$2+1$DGaugeTheory}, a mapping of a spin-$1/2$ U$(1)$ quantum link model (QLM) to an ultracold-atom system described by the Bose--Hubbard model (BHM) in $2+1$D was proposed. 
In Ref.~\cite{osborne2023spinsmathrmu1quantumlink} the BHM was extended to capture larger-spin truncations of a QLM in $1+1$D, with a concrete cold-atom proposal for $S=1$.
Here, we demonstrate how the QLM with spin $S=1$ in $2+1$D can be quantum-simulated using an ultracold-atom setup. 
Our ultracold-atom quantum simulator is described by an extended BHM that includes additional nearest-neighbor and next-nearest-neighbor interactions beyond the conventional BHM, such as can be realized by dipole--dipole interactions between magnetic atoms \cite{Dutta2015NonstandardHubbardmodels}. 
By tuning the cold-atom lattice parameters to appropriate values, the target theory emerges as an effective low-energy model at second-order perturbation theory.

This work is organized as follows.
In Section~\ref{sec:models}, we first adopt a quantum link formulation of scalar QED and embed the target theory---the \(\UN(1)\) QLM truncated to spin \(S=1\) in \(2+1\)D---into a bosonic superlattice, establishing a mapping between the local configurations of the QLM and the extended BHM.
We then propose a scheme to realize the target QLM in the cold-atom quantum simulator. 
This is accomplished by employing perturbation theory to derive an effective description of the QLM and to relate its Hamiltonian parameters to those of the extended BHM.
In Section~\ref{sec_experimental_setup}, we discuss some considerations for feasibly implementing this setup in a real cold-atom setup.
Finally, in Section~\ref{sec:numerical_results}, we validate the mapping by demonstrating the stability of gauge invariance and a high fidelity between numerical simulations of the quench dynamics of the extended BHM and those of the target QLM over all accessible evolution times.

\section{Models}
\label{sec:models}

\subsection{Quantum link model}
\label{sec:qlm}

In  the LGT of interest, matter degrees of freedom are placed on the sites of an underlying regular square lattice $\Lambda$.
In contrast, the gauge degrees of freedom reside on the links connecting these sites. 
Throughout this work, we set the lattice spacing $a$ to one.
Because the gauge-field Hilbert space is infinite-dimensional, explicit representations in quantum-simulation experiments or numerical simulations require a truncation.

To implement this truncation in $2+1$D scalar QED, we employ the QLM formulation \cite{Chandrasekharan1997QuantumLinkModels,Wiese2013UltracoldQuantumGases,Kasper2017ImplementingQuantumElectrodynamics}, in which the infinite-dimensional electric field operator $\Op{E}$, defined on the link between adjacent matter sites, is replaced by finite-dimensional spin-$S$ $z$ operators. 
Similarly, the parallel transporter $\Op{U}$, also defined on the links, is mapped to spin-$S$ ladder operators such that the canonical commutation relations between the conjugate field variables are maintained. For simplicity, and without in any way undermining the validity of our approach to go to higher-level representations of the gauge fields in two spatial dimensions, we restrict ourselves to the case of hard-core bosons for the matter fields. Consequently, we obtain the Hamiltonian \cite{Hauke2013QuantumSimulationLattice} 
\begin{align}
    \Op{H}[\qlm]=&
                    \frac{\kappa}{2\sqrt{S(S+1)}}\sum_{\vb{r},\nu}
                    \left(
                    \Op{\sigma}[\vb{r}][-]
                    \Op{s}[\vb{r},\vb{e}_{\nu}][+] 
                    \Op{\sigma}[\vb{r}+\vb{e}_{\nu}][-]
                    +\text{H.c.}
                    \right) \nonumber\\ 
                    &+\frac{\mu}{2}\sum_{\vb{r}}\Op{\sigma}[\vb{r}][z]
                    +\frac{g^2}{2}\sum_{\vb{r},\nu}  
                    \left(
                    \Op{s}[\vb{r},\vb{e}_{\nu}][z]
                    \right)^{2}
                    \ .
    \label{eq:H_QLM}
\end{align}
Here, $\vb{r}=(r_x, r_y) \in \Lambda$ denotes the coordinates of a lattice site, $\nu \in \{x,y\}$ labels the spatial direction, and $\vb{e}_{\nu}$ is the corresponding unit lattice vector.
The hard-core boson matter degrees of freedom are represented by the Pauli ladder operators $\Op{\sigma}[\vb{r}][\pm]$ acting on site $\vb{r}$, with $\Op{\sigma}[\vb{r}][z]$ related to the matter occupation number operator.
Here, $\mu$ corresponds to the bare mass.
The gauge degrees of freedom are associated with the links $(\vb{r},\vb{e}_{\nu})$ connecting adjacent sites $\vb{r}$ and $\vb{r}+\vb{e}_{\nu}$ on the lattice, and the corresponding gauge and electric fields are represented by the spin-$S$ operators $\Op{s}[\vb{r}, \vb{e}_{\nu}][+]/\sqrt{S(S+1)}$ and $\Op{s}[\vb{r}, \vb{e}_{\nu}][z]$, respectively (we scale the former by \(1/\sqrt{S(S+1)}\) to obtain the correct scaling in the large-\(S\) limit).
We note that we do not consider the plaquette operators of the gauge fields here, for simplicity, as such interactions are quite challenging to implement in our bosonic setup---they would arise as a much higher-order process due to hopping between four sites along a plaquette, but they are irrelevant for the main message of our study. It is worth noting that plaquette terms have been realized in cold-atom building-block setups \cite{Dai2017FourBodyRingExchange}.
The gauge coupling constant $g$ sets the electric energy of the model.
The hopping amplitude of the matter fields, mediated through the gauge fields, is characterized by the tunneling constant $\kappa$.
In the Hamiltonian~\eqref{eq:H_QLM}, we employ a staggered-`fermion' formulation \cite{Kogut1975HamiltonianFormulationWilsons,Susskind1977LatticeFermions}, where odd and even sites represent matter and antimatter fields, respectively. 
A particle–hole transformation has been used to remove the prefactors arising from this discretization procedure.
Further details are provided in Appendix~\ref{sec:particle_hole_transformation}.

The generator of the $\UN(1)$ gauge symmetry of the QLM acts as an analogue of Gauss’s law, and is given by
\begin{equation}
    \Op{G}[\vb{r}] = (-1)^{r_x+r_y} 
    \left[
    \frac{\Op{\sigma}[\vb{r}][z]+\mathbb{1}}{2}
    +\sum_{\nu} 
    \left(\Op{s}[\vb{r,e_{\nu}}][z]+
    \Op{s}[\vb{r-e_{\nu},e_{\nu}}][z]\right)
    \right]
    .
    \label{eq:Gauss_op}
\end{equation}
This operator commutes with the Hamiltonian, 
\begin{math}
    [\Op{H}[\qlm],\Op{G}[\vb{r}]]=0, ~ \forall \vb{r} \in \Lambda,
\end{math} 
and commutes for every different site $[\Op{G}[\vb{r}],\Op{G}[\vb{r}']]=0,~ \forall\,\vb{r}, \vb{r}', \in \Lambda$, so we may simultaneously choose the gauge sector for each site independently. 
We choose to work in the so-called physical sector (with zero background charges everywhere), which is the set of gauge-invariant states $\ket{\psi_{\text{phys}}}$ satisfying $\Op{G}[\vb{r}] \ket{\psi_{\text{phys}}} =0, ~ \forall\,\vb{r}\in\Lambda$.
We emphasize that the generator in Eq.~\eqref{eq:Gauss_op} is the truncated spin-$S$ representation of the Gauss’s law operator. 
Strictly speaking, the QLM accurately describes only the low-energy subspace of lattice QED \cite{Zohar2011ConfinementLatticeQuantumElectrodynamic,Yang2017SpindependentOpticalSuperlattice}, specifically in the regime where $\ket{S,s^{z}}$ with $s^{z} \ll S$, corresponding to the parameter range in which the
\begin{math}
    \sim g^2\sum_{\vb{r},\nu}  
    \left(\Op{s}[\vb{r},\vb{e}_{\nu}][z]\right)^{2}
\end{math}
term dominates in the Hamiltonian $\Op{H}[\qlm]$. 
In contrast, in the infinite-spin limit $S \to \infty$, the QLM correctly reproduces lattice  QED when the plaquette operator is also included. 
Furthermore, it has been demonstrated in $1+1$D that a small to moderate truncation of the bosonic gauge degrees of freedom is sufficient to approximate the untruncated QED lattice model accurately~\cite{Buyens2017FiniteRepresentationLatticeGaugeTheories,Zache2022TowardContinuumLimit,Halimeh2022achievingquantum,Lerose2024SimulatingSchwingerModel,Majcen2025OptimalLocalBasis}. 

\subsection{Bosonic mapping}
\label{sec:ultra-cold}

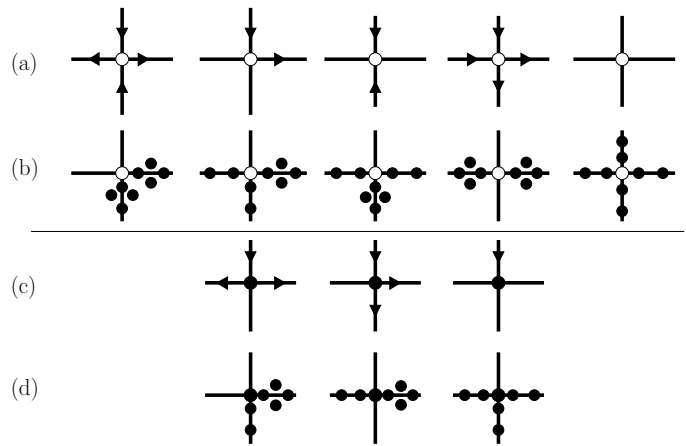
\begin{figure}[t]
    \centering
    \scalebox{0.27}{\tikzset{every picture/.style={line width=0.75pt}} 

\begin{tikzpicture}[x=0.75pt,y=0.75pt,yscale=-1,xscale=1]

\def\circrad{12} 
\def\circradb{10} 
\def\linethick{5} 
\def\arrowthick{3}

\def\ygap{15} 

\draw[line width=\linethick pt] (115,111) -- (305,111);
\draw[line width=\linethick pt] (355,111) -- (555,111);
\draw[line width=\linethick pt] (588,111) -- (778,111);
\draw[line width=\linethick pt] (818,111) -- (1008,111);
\draw[line width=\linethick pt] (1053,111) -- (1243,111);

\draw[line width=\linethick pt] (210,15) -- (210,215);
\draw[line width=\linethick pt] (450,15) -- (450,215);
\draw[line width=\linethick pt] (683,30) -- (683,200);
\draw[line width=\linethick pt] (913,30) -- (913,200);
\draw[line width=\linethick pt] (1144,30) -- (1144,200);

\filldraw[black](167,122) -- (148,111) -- (167,99) -- cycle;
\filldraw[black](240,100) -- (260,111) -- (240,123) -- cycle;
\filldraw[black](495,100) -- (515,111) -- (495,123) -- cycle;
\filldraw[black](855,100) -- (875,111) -- (855,123) -- cycle;
\filldraw[black](955,100) -- (975,111) -- (955,123) -- cycle;

\filldraw[black](200,173) -- (210,153) -- (220,173) -- cycle;
\filldraw[black](673,173) -- (683,153) -- (693,173) -- cycle;
\filldraw[black](923,152) -- (913,172) -- (903,152) -- cycle;

\filldraw[black](220,53) -- (210,73) -- (200,53) -- cycle;
\filldraw[black](460,53) -- (450,73) -- (440,53) -- cycle;
\filldraw[black](693,53) -- (683,73) -- (673,53) -- cycle;

\filldraw[black](923,53) -- (913,73) -- (903,53) -- cycle;

\draw[fill=white, fill opacity=1] (210,111) circle (\circrad);
\draw[fill=white, fill opacity=1] (450,111) circle (\circrad);
\draw[fill=white, fill opacity=1] (683,111) circle (\circrad);
\draw[fill=white, fill opacity=1] (913,111) circle (\circrad);
\draw[fill=white, fill opacity=1] (1144,111) circle (\circrad);

\begin{scope}[yshift=\ygap]
  \draw[line width=\linethick pt] (115,304) -- (305,304);
  \draw[line width=\linethick pt] (355,304) -- (555,304);
  \draw[line width=\linethick pt] (588,304) -- (778,304);
  \draw[line width=\linethick pt] (818,304) -- (1008,304);
  \draw[line width=\linethick pt] (1053,304) -- (1243,304);

  \draw[line width=\linethick pt] (210,224) -- (210,394);
  \draw[line width=\linethick pt] (450,224) -- (450,394);
  \draw[line width=\linethick pt] (683,224) -- (683,394);
  \draw[line width=\linethick pt] (913,224) -- (913,394);
  \draw[line width=\linethick pt] (1144,224) -- (1144,394);

  \draw[fill=white, fill opacity=1] (210,304) circle (\circrad);
  \draw[fill=white, fill opacity=1] (450,304) circle (\circrad);
  \draw[fill=white, fill opacity=1] (683,304) circle (\circrad);
  \draw[fill=white, fill opacity=1] (913,304) circle (\circrad);
  \draw[fill=white, fill opacity=1] (1144,304) circle (\circrad);

  \draw[fill=black, fill opacity=1] (240,304) circle (\circradb);
  \draw[fill=black, fill opacity=1] (264,286) circle (\circradb);
  \draw[fill=black, fill opacity=1] (264,322) circle (\circradb);
  \draw[fill=black, fill opacity=1] (288,304) circle (\circradb);
  
  \draw[fill=black, fill opacity=1] (374,304) circle (\circradb);
  \draw[fill=black, fill opacity=1] (418,304) circle (\circradb);
  
  \draw[fill=black, fill opacity=1] (482,304) circle (\circradb);
  \draw[fill=black, fill opacity=1] (508,286) circle (\circradb);
  \draw[fill=black, fill opacity=1] (508,322) circle (\circradb);
  \draw[fill=black, fill opacity=1] (534,304) circle (\circradb);
  
  \draw[fill=black, fill opacity=1] (610,304) circle (\circradb);
  \draw[fill=black, fill opacity=1] (655,304) circle (\circradb);
  
  \draw[fill=black, fill opacity=1] (715,304) circle (\circradb);
  \draw[fill=black, fill opacity=1] (760,304) circle (\circradb);
  
  \draw[fill=black, fill opacity=1] (840,304) circle (\circradb);
  \draw[fill=black, fill opacity=1] (860,284) circle (\circradb);
  \draw[fill=black, fill opacity=1] (860,324) circle (\circradb);
  \draw[fill=black, fill opacity=1] (880,304) circle (\circradb);

  \draw[fill=black, fill opacity=1] (945,304) circle (\circradb);
  \draw[fill=black, fill opacity=1] (965,284) circle (\circradb);
  \draw[fill=black, fill opacity=1] (965,324) circle (\circradb);
  \draw[fill=black, fill opacity=1] (985,304) circle (\circradb);

  \draw[fill=black, fill opacity=1] (1075,304) circle (\circradb);
  \draw[fill=black, fill opacity=1] (1115,304) circle (\circradb);
  
  \draw[fill=black, fill opacity=1] (1175,304) circle (\circradb);
  \draw[fill=black, fill opacity=1] (1220,304) circle (\circradb);

  \draw[fill=black, fill opacity=1] (210,330) circle (\circradb);
  \draw[fill=black, fill opacity=1] (190,345) circle (\circradb);
  \draw[fill=black, fill opacity=1] (230,345) circle (\circradb);
  \draw[fill=black, fill opacity=1] (210,370) circle (\circradb);
  
  \draw[fill=black, fill opacity=1] (450,330) circle (\circradb);
  \draw[fill=black, fill opacity=1] (450,370) circle (\circradb);
  
  \draw[fill=black, fill opacity=1] (683,330) circle (\circradb);
  \draw[fill=black, fill opacity=1] (665,349) circle (\circradb);
  \draw[fill=black, fill opacity=1] (701,349) circle (\circradb);
  \draw[fill=black, fill opacity=1] (683,370) circle (\circradb);
  
  \draw[fill=black, fill opacity=1] (1144,245) circle (\circradb);
  \draw[fill=black, fill opacity=1] (1144,275) circle (\circradb);
  
  \draw[fill=black, fill opacity=1] (1144,335) circle (\circradb);
  \draw[fill=black, fill opacity=1] (1144,375) circle (\circradb);
\end{scope}

\draw (40,433) -- (1260,433);


\draw[line width=\linethick pt] (365,529) -- (535,529);
\draw[line width=\linethick pt] (598,529) -- (768,529);
\draw[line width=\linethick pt] (828,529) -- (998,529);

\draw[line width=\linethick pt] (450,449) -- (450,620);
\draw[line width=\linethick pt] (683,449) -- (683,620);
\draw[line width=\linethick pt] (913,449) -- (913,620);

\filldraw[black](407,541) -- (387,529) -- (407,518) -- cycle;
\filldraw[black](495,518) -- (515,529) -- (495,541) -- cycle;
\filldraw[black](710,518) -- (730,529) -- (710,541) -- cycle;

\filldraw[black](460,471) -- (450,491) -- (440,471) -- cycle;
\filldraw[black](693,471) -- (683,491) -- (673,471) -- cycle;
\filldraw[black](923,471) -- (913,491) -- (903,471) -- cycle;
\filldraw[black](693,571) -- (683,591) -- (673,571) -- cycle;

\draw[fill=black, fill opacity=1] (450,529) circle (\circrad);
\draw[fill=black, fill opacity=1] (683,529) circle (\circrad);
\draw[fill=black, fill opacity=1] (913,529) circle (\circrad);

\begin{scope}[yshift=\ygap]
\draw[line width=\linethick pt] (365,719) -- (535,719);
\draw[line width=\linethick pt] (598,719) -- (768,719);
\draw[line width=\linethick pt] (828,719) -- (998,719);

\draw[line width=\linethick pt] (450,639) -- (450,810);
\draw[line width=\linethick pt] (683,639) -- (683,810);
\draw[line width=\linethick pt] (913,639) -- (913,810);

\draw[fill=black, fill opacity=1] (450,719) circle (\circrad);
\draw[fill=black, fill opacity=1] (683,719) circle (\circrad);
\draw[fill=black, fill opacity=1] (913,719) circle (\circrad);

\draw[fill=black, fill opacity=1] (474,719) circle (\circradb);
\draw[fill=black, fill opacity=1] (497,700) circle (\circradb);
\draw[fill=black, fill opacity=1] (497,738) circle (\circradb);
\draw[fill=black, fill opacity=1] (520,719) circle (\circradb);

\draw[fill=black, fill opacity=1] (620,719) circle (\circradb);
\draw[fill=black, fill opacity=1] (655,719) circle (\circradb);

\draw[fill=black, fill opacity=1] (707,719) circle (\circradb);
\draw[fill=black, fill opacity=1] (731,702) circle (\circradb);
\draw[fill=black, fill opacity=1] (731,737) circle (\circradb);
\draw[fill=black, fill opacity=1] (753,719) circle (\circradb);

\draw[fill=black, fill opacity=1] (850,719) circle (\circradb);
\draw[fill=black, fill opacity=1] (885,719) circle (\circradb);

\draw[fill=black, fill opacity=1] (940,719) circle (\circradb);
\draw[fill=black, fill opacity=1] (980,719) circle (\circradb);

\draw[fill=black, fill opacity=1] (450,744) circle (\circradb);
\draw[fill=black, fill opacity=1] (450,784) circle (\circradb);
  
\draw[fill=black, fill opacity=1] (913,744) circle (\circradb);
\draw[fill=black, fill opacity=1] (913,784) circle (\circradb);
\end{scope}

\def\textscale{1.3}
\draw (0,97) node[scale=\textscale] [anchor=north west][inner sep=0.75pt] [align=left] {{\Huge (a)}};
\draw (0,293) node[scale=\textscale] [anchor=north west][inner sep=0.75pt] [align=left] {{\Huge (b)}};
\draw (0,513) node[scale=\textscale] [anchor=north west][inner sep=0.75pt] [align=left] {{\Huge (c)}};
\draw (0,705) node[scale=\textscale] [anchor=north west][inner sep=0.75pt] [align=left] {{\Huge (d)}};

\end{tikzpicture}}
    \caption{Selection of Gauss’s-law-allowed matter and electric-field configurations in the QLM and their corresponding representations in the BHM. 
    Left- and downward-pointing arrows indicate electric-flux operator eigenvalues $\Op{s}[\vb{r}][z] = -1$, right- and upward-pointing arrows correspond to $\Op{s}[\vb{r}][z] = 1$,  and links without arrows denote $\Op{s}[\vb{r}][z] = 0$.
    (a,b)~QLM configurations with unoccupied matter sites and their bosonic counterparts. 
    (c,d) QLM configurations with an occupied matter site and the corresponding bosonic representation.
    All other configurations consistent with Gauss’s law can be generated by applying rotations and reflections to these vertices.
    }
    \label{fig:gauge_states_mapping}
\end{figure}
To perform a mapping of the allowed local configurations of the QLM to a bosonic system, we embed the QLM in an optical superlattice \cite{Jaksch1998ColdBosonicAtoms} of size $ L\times L$, where $L$ is an even integer equal to twice the linear extent of the QLM lattice in units of the $N$ matter sites along one direction.
We denote sites of the bosonic lattice by $\vb{j}=(j_x, j_y)$ (while we use \(\vb{r}\) to represent matter sites in the QLM).
Matter sites are located at positions where both $j_x$ and $j_y$ are even integers and correspond to QLM matter sites labelled by $\vb{r}=(j_x/2,\, j_y/2)$.
The gauge sites are located at positions where exactly one of $j_x$ or $j_y$ is even, representing the gauge site in the QLM on the link \((\vb{r}, \vb{e}_x)\), \(\vb{r} = ([j_x-1]/2, j_y/2)\), when \(j_x\) is odd, or \((\vb{r}, \vb{e}_y)\), \(\vb{r} = (j_x/2, [j_y-1]/2)\), when \(j_y\) is odd.
Sites where both $j_x$ and $j_y$ are odd are designated as forbidden and are removed from the superlattice, since the corresponding positions do not map to matter or gauge degrees of freedom of the QLM.
We define local bosonic ladder operators $\Op{b}[\vb{j}]$ and $\Op{b}[\vb{j}]*$ on each site, which satisfy the canonical commutation relations $[\Op{b}[\vb{j}], \Op{b}[\vb{j}']] = [\Op{b}[\vb{j}]*, \Op{b}[\vb{j}']*] = 0$ and $[\Op{b}[\vb{j}], \Op{b}[\vb{j}']*] = \delta_{\vb{j},\vb{j}'}$ for all $\vb{j}, \vb{j}'$.
The number operator is defined by $\Op{n}[\vb{j}] = \Op{b}[\vb{j}]*\, \Op{b}[\vb{j}]$.

At matter sites, the local Hilbert space $\mathcal{H}_m$ is restricted to bosonic occupation numbers $n_m \in \{0,1\}$ (hard-core constraint), which correspond to unoccupied and occupied matter sites, respectively, in the particle–hole transformed QLM.
On gauge sites, the local Hilbert space $\mathcal{H}_g$ is restricted to even bosonic occupation numbers $n_g \in \{0,2,\ldots,4S\}$. 
The eigenvalues $\{-S,-S+1,\ldots,S\}$ of the electric flux operator $ \Op{s}[\vb{r},\vb{e}_{\nu}][z]$ are mapped to local bosonic occupations $\{0,2,\ldots,4S\}$, respectively.
The mapping of local configurations is chosen such that gauge-invariant configurations are connected through second-order processes in the Bose--Hubbard simulator.
In Fig.~\ref{fig:gauge_states_mapping}, we present some of the configurations of matter sites and their surrounding gauge links that are allowed by Gauss’s law, together with their corresponding bosonic representations. 
The local operators of the QLM Hamiltonian can be written in terms of bosonic creation and annihilation operators (which we label in terms of their mapped indices in the QLM), projected onto the constrained local Hilbert spaces as
\begin{subequations}\label{eq:projection_ops}
\begin{align}
    &\Op{\sigma}[\vb{r}][+]=
    \Op{P}[\vb{r}] 
    \Op{b}_{\vb{r}}^{\dagger} \Op{P}[\vb{r}]*, \qquad 
    \Op{\sigma}[\vb{r}][z]=\Op{P}[\vb{r}] 
    \left(2 \Op{n}[\vb{r}]-1\right) \Op{P}[\vb{r}]*,
    \\
    &\Op{\tau}[\vb{r}, \vb{e}_{\nu}][+]=
    \frac{1}{\sqrt{2S(2S+1)}} 
    \Op{P}[\vb{r}, \vb{e}_{\nu}] 
    \left(\Op{b}[\vb{r}, \vb{e}_{\nu}]*\right)^{2}
    \Op{P}[\vb{r}, \vb{e}_{\nu}]*,\label{eq:tau}
    \\
    &\Op{s}[\vb{r}, \vb{e}_{\nu}][z]=
    \frac{1}{2S} \Op{P}[\vb{r}, \vb{e}_{\nu}]
    \left(\Op{n}[\vb{r},\vb{e}_{\nu}]-2S\right)
    \Op{P}[\vb{r},\vb{e}_{\nu}]*
    \ .
\end{align}
\end{subequations}
Here, $\Op{P}[\vb{r}]$ is a projection onto the matter Hilbert space $\mathcal{H}_m$ attached to site $\vb{r}$, while $\Op{P}[\vb{r}, \vb{e}_{\nu}]$ projects onto the gauge Hilbert space $\mathcal{H}_{g}$ associated with link $(\vb{r}, \vb{e}_{\nu})$. 
We note that in our mapping, a ladder operator \(\hat{\tau}^+\)~\eqref{eq:tau} arises that differs from the conventional spin raising operator \(\hat{s}^+\) that is usually used in the quantum link formulation.
Nevertheless, as discussed in Ref.~\cite{osborne2023spinsmathrmu1quantumlink}, this ladder operator reproduces the correct scaling in the large-\(S\) limit, and the results only have minor quantitative differences, with similar qualitative behavior being retained.
Therefore, in the remainder of this work, we shall substitute \(\hat{\tau}^+\) for the conventional \(\hat{s}^+\) operator in our formulation of the QLM. 

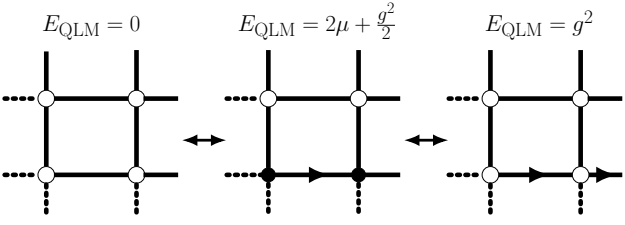
\begin{figure}[t]
    \centering
    \scalebox{0.35}{\tikzset{every picture/.style={line width=0.75pt}} 
\begin{tikzpicture}[x=0.75pt,y=0.75pt,yscale=-1,xscale=1]
\def\circrad{12} 
\def\circradb{10} 

\def\linethick{5} 

\draw[line width=\linethick pt](151,120) -- (258,120) ;
\draw[line width=\linethick pt](280,120) -- (330,120) ;
\draw[loosely dashed,line width=\linethick pt,line cap=round]    (80,120) -- (127,120) ;
 
\draw[line width=\linethick pt](151,230) -- (258,230) ;
\draw[line width=\linethick pt](280,230) -- (330,230) ;
\draw[loosely dashed,line width=\linethick pt,line cap=round]    (80,230) -- (127,230) ;

\draw[line width=\linethick pt](472,120) -- (578,120) ;
\draw[line width=\linethick pt](602,120) -- (650,120) ;
\draw[loosely dashed,line width=\linethick pt,line cap=round]    (400,120) -- (447,120) ;


\draw[
  line width=\linethick pt,
  postaction={decorate},
  decoration={
    markings,
    mark=at position 0.45 with {\arrow[scale=1.6]{Latex[length=5mm, width=4mm]}}
  }
] (466,230) -- (650,230);

\draw[loosely dashed,line width=\linethick pt,line cap=round]    (400,230) -- (466,230) ;

\draw[line width=\linethick pt](792,120) -- (898,120) ;
\draw[line width=\linethick pt](920,120) -- (970,120) ;
\draw[loosely dashed,line width=\linethick pt,line cap=round]    (720,120) -- (767,120) ;

\draw[mid arrow,line width=\linethick pt](792,230) -- (898,230) ;
\draw[out arrow,line width=\linethick pt](922,230) -- (970,230) ;
\draw[loosely dashed,line width=\linethick pt,line cap=round]    (720,230) -- (767,230) ;


\draw[line width=\linethick pt](140,52) -- (140,108) ;
\draw[line width=\linethick pt](140,132) -- (140,218) ;
\draw[loosely dashed,line width=\linethick pt,line cap=round] (140,244) -- (140,290) ;

\draw[line width=\linethick pt](270,50) -- (270,108) ;
\draw[line width=\linethick pt](270,132) -- (270,218) ;
\draw[loosely dashed,line width=\linethick pt,line cap=round] (270,244) -- (270,290) ;

\draw[line width=\linethick pt](460,50) -- (460,108) ;
\draw[line width=\linethick pt](460,132) -- (460,220) ;
\draw[loosely dashed,line width=\linethick pt,line cap=round] (460,243) -- (460,284) ;

\draw[line width=\linethick pt](590,50) -- (590,108) ;
\draw[line width=\linethick pt](590,132) -- (590,220) ;
\draw[loosely dashed,line width=\linethick pt,line cap=round] (590,243) -- (590,284) ;

\draw[line width=\linethick pt](780,50) -- (780,108) ;
\draw[line width=\linethick pt](780,132) -- (780,218) ;
\draw[loosely dashed,line width=\linethick pt,line cap=round] (780,244) -- (780,290) ;

\draw[line width=\linethick pt](910,50) -- (910,108) ;
\draw[line width=\linethick pt](910,132) -- (910,218) ;
\draw[loosely dashed,line width=\linethick pt,line cap=round] (910,244) -- (910,290) ;

\draw [black, line width=1] (270,230) circle (\circrad);
\draw [black, line width=1] (140,120) circle (\circrad);
\draw [black, line width=1] (270,120) circle (\circrad);
\draw [black, line width=1] (140,230) circle (\circrad);

\draw [fill=black] (460,230) circle (\circradb);
\draw [fill=black] (590,230) circle (\circradb);

\draw [black, line width=1] (590,120) circle (\circrad);
\draw [black, line width=1] (460,120) circle (\circrad);

\draw [black, line width=1] (780,120) circle (\circrad);
\draw [black, line width=1] (780,230) circle (\circrad);
\draw [black, line width=1] (910,230) circle (\circrad);
\draw [black, line width=1] (910,120) circle (\circrad);

\node[anchor=mid] at (205,20) {\Huge$E_{\mathrm{QLM}} = 0$};
\node[anchor=mid] at (530,20) {\Huge$E_{\mathrm{QLM}} = 2\mu+\frac{g^2}{2}$};
\node[anchor=mid] at (850,20) {\Huge$E_{\mathrm{QLM}} = g^2$};

\draw[line width=3pt, {Latex[length=6mm]}-{Latex[length=6mm]}] (335,180) -- (400,180);
\draw[line width=3pt, {Latex[length=6mm]}-{Latex[length=6mm]}] (655,180) -- (720,180);

\end{tikzpicture}}
    \caption{Configurations of the QLM on the unit cell that are connected via second-order processes in the extended BHM. 
    To restrict the quantum simulator to the gauge-invariant subspace, these states are tuned into mutual resonance.
    }
    \label{fig:states_resonance}
\end{figure}
Focusing henceforth on the case of $S=1$, this mapping can be realized in an ultracold-atom setup described by the extended Bose--Hubbard Hamiltonian
\begin{align}
    \Op{H}[\bhm]=
    &-J \sum_{\vb{j},\nu} 
    \left(\Op{b}[\vb{j}]* \Op{b}_{\vb{j}+\vb{e}_{\nu}}+\mathrm{H.c.}\right)
    +\sum_{\vb{j}=1} \frac{U_{\vb{j}}}{2} 
    \,\Op{n}_{\vb{j}}\left(\Op{n}_{\vb{j}}-1\right)
    \nonumber \\
    &+\sum_{\vb{j}} 
    \left(\boldsymbol{\gamma}\cdot\vb{j}-\delta_{\vb{j}}\right) \Op{n}_{\vb{j}}
    + V \sum_{\vb{j},\nu}
    \Op{n}_{\vb{j}} \Op{n}_{\vb{j}+\vb{e}_{\nu}}
    \nonumber\\
    &+W \sum_{\vb{j}, \{\nu,\nu'\} \in \Omega}
    \Op{n}_{2\vb{j}+\vb{e}_{\nu}} \Op{n}_{2\vb{j}+\vb{e}_{\nu'}}.
    \label{eq:H_BHM}
\end{align}
The tunneling amplitude is denoted by $J$, while the on-site interaction strength $U_j$ equals $\alpha\,U$ on matter sites and $U$ on gauge sites. 
Here, $\alpha$ denotes a tunable ratio of on-site interactions, which can be adjusted experimentally through differences in the Wannier functions of two Bloch bands
\cite{Jaksch1998ColdBosonicAtoms,Bloch2008ManyBodyPhysics,Olschlager2013InteractioninducedChiralPxpmipy}.
A tilt vector $\boldsymbol{\gamma} = (\gamma_{x}, \gamma_{y})$ encodes the potential gradient along each spatial direction and stabilizes $\UN(1)$ gauge-invariant configurations \cite{Halimeh2020RobustnessOfGaugeInvariantDynamics,Halimeh2021GaugeSymmetryProtection,Facchi2002QuantumZenoSubspaces,Facchi2004UnificationDynamicalCoupling,Facchi2009QuantumZenoDynamics,Abanin2017RigorousTheoryManyBody,Burgarth2019generalized}.
A chemical potential $\delta_{\vb{j}}$ applied only on gauge links (zero on matter sites) generates a staggered potential with deeper gauge sites and shallower matter sites, which allows us to ensure the correct processes are resonant. 
Furthermore, assuming magnetic atoms with dipole--dipole interactions (see next section), we include nearest-neighbor interactions with strength $V$ and next-nearest-neighbor interactions with strength $W$ between two consecutive \emph{gauge} sites (i.e.\ gauge sites which are both adjacent to the same matter site), with \(\Omega\) denoting set of all two-element subsets of \(\{+x, +y, -x, -y\}\).
These interactions ensure that local configurations satisfying Gauss’s law have the same energy in the extended BHM if and only if they have the same energy in the original QLM formulation.
The Gauss’s law operator can then be rewritten in terms of bosonic operators in the number operator basis as
\begin{equation}
    \Op{G}[\vb{j}]=(-1)^{(j_x+j_y)/2}
                        \left[\Op{n}_{\vb{j}}
                        +\frac{1}{2}\sum_{\nu}\left( \Op{n}_{\vb{j}+\vb{e}_{\nu}}
                        +\Op{n}_{\vb{j}-\vb{e}_{\nu}}
                        -4\right)
                        \right],
\end{equation}
where $\vb{j}$ is the index of a matter site (i.e.\ with both elements even).
Working in the regime $U \sim 2\delta \gg J$, the model can be treated perturbatively in the hopping amplitude~$J$, which provides the only off-diagonal contributions to the Hamiltonian in the bosonic occupation-number basis.
The diagonal part of the extended Bose--Hubbard model can then be written as
\begin{align}
    \Op{H}[\mathrm{diag}]=&\sum_{\vb{j}} 
    \frac{U_{\vb{j}}}{2} 
    \Op{n}_{\vb{j}}\left(\Op{n}_{\vb{j}}-1\right)
    -\sum_{\vb{j}} 
    \delta_{\vb{j}} \Op{n}_{\vb{j}} 
    + V \sum_{\vb{j},\nu}
    \Op{n}_{\vb{j}} \Op{n}_{\vb{j}+\vb{e}_{\nu}}
    \nonumber\\
    &+W \sum_{\vb{j},\{\nu,\nu'\}\in \Omega}
    \Op{n}_{2\vb{j}+\vb{e}_{\nu}} \Op{n}_{2\vb{j}+\vb{e}_{\nu'}} 
    +\Op{H}[G].
\end{align}
Here, we have identified a linear gauge protection term
\begin{math}
    \Op{H}[G]=2\sum_{\vb{j}\in \text{matter}}  (-1)^{(j_x+j_y)/2}\,\boldsymbol{\gamma}\cdot\vb{j} 
    \, \Op{G}_{\vb{j}},
\end{math}
which acts as a global energy-penalty term, effectively projecting the extended BHM onto the gauge-invariant subspace during the time evolution.
In Ref.~\cite{Halimeh2021GaugeSymmetryProtection}, it was shown---using exact numerics on finite-size systems---that for an arbitrary initial state $\ket{\psi}$ within the target symmetry sector, such linear gauge protection terms can keep the gauge violation throughout quench dynamics at a perturbatively small level for arbitrarily long times. 

The mapping between the QLM and the extended BHM is obtained by relating the parameters of the QLM Hamiltonian~\eqref{eq:H_QLM} to those of the extended BHM Hamiltonian~\eqref{eq:H_BHM} through second-order perturbation theory, as detailed in Appendix~\ref{sec:perturbation_theory}.
Within this framework, second-order processes connect bosonic configurations that satisfy Gauss’s law, and bringing these configurations into resonance imposes two constraints on the BHM parameters.
To derive these constraints, we introduce translationally invariant unit cells corresponding to a single QLM plaquette with periodic boundary conditions (PBC). 
In the BHM, each such unit cell maps onto a $4 \times 4$ lattice of bosonic sites with PBC (see, e.g., Fig.~\ref{fig:states_resonance}).
This procedure leads to the following set of equations
\begin{subequations}
\begin{align}
    5U - 12 V + 16 W - 2 \delta &\approx 0 \label{eq:resonance_A},\\
    U - 12V + 24W - 2 \delta &\approx 0 \label{eq:resonance_B}.
\end{align}
\end{subequations}
Only the states depicted in Fig.~\ref{fig:states_resonance} are required to derive these relations, as additional states connected by second-order processes within the unit cell yield equivalent constraints.

In the cold-atom quantum simulator, additional unwanted second-order processes may occur. 
Specifically, starting from configurations that satisfy Gauss’s law, the system may transition to a state without a native counterpart in the QLM and then return to the initial configuration.
In Fig.~\ref{fig:state_table}, some of these processes are illustrated, in particular those that either originate from configurations that involve doublons on a neighboring site [Fig.~\ref{fig:state_table}(a–c)] or those with only a single boson [Fig.~\ref{fig:state_table}(d)] in a matter-gauge pair of sites.
Within the $S=1$ truncation, some configurations contain gauge sites with four bosons on the gauge site.
The processes illustrated in Fig.~\ref{fig:state_table} contribute to the renormalization of the mass parameter $\mu$, among other processes not shown.
We note that these expressions closely resemble those obtained in the $2+1$D case for an $S=\tfrac{1}{2}$ truncation \cite{Osborne2025Scale$2+1$DGaugeTheory}. 
In particular, upon removing the next-neighbor interaction of strength $V$ and the gauge–gauge interaction of strength $W$, the expressions coincide up to a combinatorial factor in the numerator.

Furthermore, the presence of long-range interactions in the BHM can result in unwanted interactions appearing in the mapped QLM, which can be removed by imposing \(V = 2W\), as in Ref.~\cite{osborne2023spinsmathrmu1quantumlink}.
Using this relation in addition to the two resonance conditions~\eqref{eq:resonance_A} and \eqref{eq:resonance_B}, the six parameters of the BHM are reduced to three, matching the number of parameters in the target QLM. 
Owing to the tilt $\gamma$ in Eq.~\eqref{eq:H_BHM}, different gradient potentials are applied along the $x$ and $y$ directions of the lattice, thereby suppressing unwanted resonances.
As a result, the expressions for the QLM parameters corresponding to the Bose--Hubbard quantum simulator will depend on the direction, but as the tilt~\(\gamma\) is relatively small compared to the other couplings in the BHM, this directionality is expected to be negligible, and we can just use the average of the expressions in the \(x\) and \(y\) directions.
The mapping between the QLM coefficients and those of the extended BHM is as follows (for a detailed derivation, see Appendix~\ref{sec:perturbation_theory}):
\begin{widetext}
\begin{subequations}
\begin{align}
    \kappa_{i}=& 
    \sqrt{6}J^{2}\left( \frac{1}{5 V - 12 W + \delta + \gamma_{i}}
    + (\gamma_{i} \rightarrow -\gamma_{i}) \right) 
    +\sqrt{6}J^{2} \left(
    \frac{1}{U - 7 V + 12 W - \delta + \gamma_{i}} 
    + (\gamma_{i} \rightarrow -\gamma_{i}) \right),
    \\
    g_{i}^2=&
    8 J^{2} \left(
    \frac{1}{3 U - 7 V + 8 W - \delta + \gamma_{i}} 
    + (\gamma_{i} \rightarrow -\gamma_{i})
    \right)
    -8 J^{2}\left( 
    \frac{1}{U - 7 V + 12 W - \delta + \gamma_{i}} 
    +(\gamma_{i} \rightarrow -\gamma_{i})
    \right)
    +8 U - 16 W ,\\\nonumber
    \mu_{i}=& 
    +4 J^{2}\left(
    \frac{1}{(1- \alpha)U - 3 V + 8 W - \delta + \gamma_{j}} 
    + (\gamma_{j} \rightarrow -\gamma_{j}) 
    \right)
    +2 J^{2}\left( 
    \frac{1}{(1- \alpha)U - 3 V + 8 W - \delta + \gamma_{i}} 
    + (\gamma_{i} \rightarrow -\gamma_{i}) 
    \right)
    \nonumber\\
    &
    -J^{2}\left( 
    \frac{1}{3 U - 7 V + 8 W - \delta + \gamma_{i}} 
    + (\gamma_{i} \rightarrow -\gamma_{i}) 
    \right)
    -2 J^{2}\left( 
    \frac{1}{U - 7 V + 12 W - \delta + \gamma_{j}} 
    + (\gamma_{j} \rightarrow -\gamma_{j}) 
    \right)
    \nonumber\\
    &
    -J^{2}\left( 
    \frac{1}{U - 7 V + 12 W - \delta + \gamma_{i}} 
    + (\gamma_{i} \rightarrow -\gamma_{i}) 
    \right)
    +3 J^{2}\left( 
    \frac{1}{- 2 U + 5 V - 8 W + \delta + \gamma_{j}} 
    + (\gamma_{j} \rightarrow -\gamma_{j}) 
    \right)
    \nonumber\\
    &
    + \frac{3 J^{2}}{2}\left(
    \frac{1}{- 2 U + 5 V - 8 W + \delta + \gamma_{i}} 
    + (\gamma_{i} \rightarrow -\gamma_{i}) 
    \right)
    +\frac{J^{2}}{2}\left( 
    \frac{1}{5 V - 12 W + \delta + \gamma_{i}} 
    + (\gamma_{i} \rightarrow -\gamma_{i}) 
    \right)
    \nonumber\\
    &- \frac{3 U}{2} + 6 V - 10 W + \delta,
\end{align}
\end{subequations}
\end{widetext}
where $i,j\in \{x,y\}$ are the directions with the lattice $\Lambda$, and \((\gamma_{i} \rightarrow -\gamma_{i})\) denotes repeating the previous expression but with \(\gamma_{i}\) replaced by \(-\gamma_{i}\).

\begin{figure}[t]
\centering
\scalebox{0.94}{$
\begin{array}{l c c c c l}
\text{(a)} & \well{1}{2}  & \leftrightarrow & \well{2}{1} &  & \displaystyle\sum_{\pm \gamma_i} \frac{16J^2}{(1-\alpha)U - 3V +8W -\delta +\gamma_{i}}\\[10pt]
\text{(b)} & \well{1}{2}  & \leftrightarrow & \well{0}{3} &  & \displaystyle\sum_{\pm \gamma_i} \frac{12 J^2}{-2U + 5V -8W +\delta +\gamma_{i}} \\[10pt]
\text{(c)} & \well{0}{2}  & \leftrightarrow & \well{1}{1} &  & \displaystyle\sum_{\pm \gamma_i} \frac{8 J^2}{U - 7V +12W -\delta +\gamma_{i}} \\[10pt]
\text{(d)} & \well{1}{0}  & \leftrightarrow & \well{0}{1} &  & \displaystyle\sum_{\pm \gamma_i} \frac{2 J^2}{ 5V -12W +\delta +\gamma_{i}}\\[10pt]
\end{array}
$}
\caption{ 
    Undesired second-order processes in the mapping of the $2 + 1$D spin-$1$ $\UN(1)$ QLM onto a two-dimensional Bose--Hubbard optical superlattice, where the shallow well represents a matter site, and the deep well one of its neighboring gauge sites. These are undesired as the configurations on the right do not correspond to any valid QLM configuration. On the right, we show the corresponding perturbative contributions to the matrix elements, demonstrating that these processes renormalize the mass parameter $\mu$. The list is not exhaustive.}
\label{fig:state_table}
\end{figure}

\section{Experimental setup}
\label{sec_experimental_setup}

\begin{figure}[t]
    \centering
    \scalebox{1.15}{

\begin{tikzpicture}[
  line cap=round,
  line join=round,
  x={(1.15cm,0cm)},
  y={(0.75cm,0.35cm)},
  z={(0cm,1.05cm)}
]

\def\a{1}
\def\b{1.3}
\def\sh{0.}

\tikzset{
  link/.style={line width=0.9pt, draw=black},
  inter/.style={line width=0.8pt, draw=gray!70, dashed},
  site/.style={circle, inner sep=0pt, minimum size=7pt, fill=red!70!black, draw=white, line width=0.4pt},
  arrowmark/.style={-Stealth, line width=0.8pt}
}

\foreach \i in {0,1,2,3}{
  \foreach \j in {0,1,2}{
    \coordinate (g-\i-\j) at ({\i*\a},{\j*\a},0);
  }
}

\foreach \i in {0,1,2,3}{
  \foreach \j in {0,1,2}{
    \coordinate (m-\i-\j) at ({(\i+\sh)*\a},{\j*\a},\b);
  }
}

\foreach \j in {0,1,2}{
  \foreach \i [evaluate=\i as \ip using int(\i+1)] in {0,1,2}{
    \draw[link] (g-\i-\j) -- (g-\ip-\j);
  }
}
\foreach \i in {0,1,2,3}{
  \foreach \j [evaluate=\j as \jp using int(\j+1)] in {0,1}{
    \draw[link] (g-\i-\j) -- (g-\i-\jp);
  }
}

\foreach \j in {0,1,2}{
  \foreach \i [evaluate=\i as \ip using int(\i+1)] in {0,1,2}{
    \coordinate (gh-\i-\j) at ($(g-\i-\j)!0.5!(g-\ip-\j)$);
    \node[site] at (gh-\i-\j) {}; 
  }
}
\foreach \i in {0,1,2,3}{
  \foreach \j [evaluate=\j as \jp using int(\j+1)] in {0,1}{
    \coordinate (gv-\i-\j) at ($(g-\i-\j)!0.5!(g-\i-\jp)$);
    \node[site] at (gv-\i-\j) {}; 
  }
}

\foreach \j in {0,1,2}{
  \foreach \i [evaluate=\i as \ip using int(\i+1)] in {0,1,2}{
    \draw[link] (m-\i-\j) -- (m-\ip-\j);
  }
}
\foreach \i in {0,1,2,3}{
  \foreach \j [evaluate=\j as \jp using int(\j+1)] in {0,1}{
    \draw[link] (m-\i-\j) -- (m-\i-\jp);
  }
}

\foreach \i in {0,1,2,3}{
  \foreach \j in {0,1,2}{
    \draw[inter] (g-\i-\j) -- (m-\i-\j);
  }
}

\foreach \i in {0,1,2,3}{
  \foreach \j in {0,1,2}{
    \draw[arrowmark, draw=orange!120]
      (m-\i-\j) ++(0.28,-0.8,0) -- ++(0.85,-0.5,0.73);
  }
}

\foreach \j in {0,1,2}{
  \foreach \i in {0,1,2}{
    \draw[arrowmark, draw=orange!120] (gh-\i-\j) ++(0,0,-0.3) -- ++(0,0,0.7);
  }
}
\foreach \i in {0,1,2,3}{
  \foreach \j in {0,1}{
    \draw[arrowmark, draw=orange!120] (gv-\i-\j)++(0,0,-0.3) -- ++(0,0,0.7);
  }
}

\foreach \i in {0,1,2,3}{
  \foreach \j in {0,1,2}{
    \node[site] at (m-\i-\j) {};
  }
}

\node[anchor=west, font=\large] at ($(m-3-2)+(-0.9,0,0.6)$) {Matter};
\node[anchor=west, font=\large, yshift=-5pt] at ($(g-3-0)+(0.45,0,0)$) {Gauge};

\coordinate (O) at (-0.4,-0.7,-0.3);
\draw[->, thick, gray!90] (O) -- ++(1.1,0,0) node[below right] {$x$};
\draw[->, thick, gray!90] (O) -- ++(0,1.1,0) node[left,xshift=-4pt ,yshift=3pt] {$y$};
\draw[->, thick, gray!90] (O) -- ++(0,0,1.1) node[above] {$z$};

\begin{scope}[transform canvas={yshift=-12pt}]
  \draw[decorate, decoration={brace, mirror, amplitude=5pt}, thick]
    (g-2-0) -- (g-3-0)
    node[midway, below=6pt] {$a$};
\end{scope}

\begin{scope}[transform canvas={xshift=-7pt}]
  \draw[decorate, decoration={brace, amplitude=5pt}, thick]
    (g-0-0) -- ++(0,0,\b)
    node[midway, left=6pt] {$h$};
\end{scope}

\end{tikzpicture}}
    \caption{Schematic diagram of the proposed cold-atom simulator architecture. The upper layer
    hosts the matter sites, while the lower layer contains the gauge and forbidden sites.
    The polarisation direction in the upper layer is chosen to be
    $d=\frac{1}{\sqrt{3}}(1,1,1)$.
        }
    \label{fig:scheme_simulator}
\end{figure}
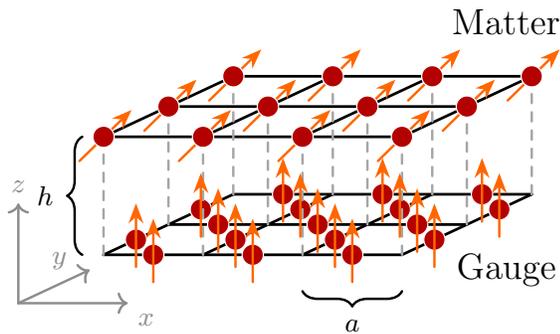

In this section, we propose an architecture for realizing the simulator based on two layers of superlattices.
The two superlattices are arranged parallel to the $x$–$y$ plane, with the lower layer at
$z=0$ and the upper layer at $z=h$, where $h$ is the interlayer separation.
Relative to one another, the layers are shifted by $a/2$ along both the $x$ and $y$
directions, where $a$ is the lattice spacing. 
By convention, we choose that the upper layer hosts the matter sites, while the lower layer contains the gauge and forbidden sites. 
The architecture is shown in Fig.~\ref{fig:scheme_simulator}.
We arrange the sites in this fashion in order to use a magnetic gradient field to rotate the matter dipoles to the so-called magic angles, eliminating unwanted leading-order long-range interactions, as we shall discuss below.
Within each layer, we apply a bichromatic superlattice potential 
\begin{equation}
    V_{m,g}
    = \sum_{r=x,y}\Big[ V_s\,\cos^2(4\pi r/\lambda)
    - V_l\,\cos^2(2\pi r/\lambda-\phi_{m,g}) \Big],
\end{equation}
where $\phi_{m,g}$ is a layer dependent phase shift with $\phi_{g}=\frac{\pi}{4}$ and $\phi_{m}=\frac{3\pi}{4}$, $(x,y)$ denotes the position within the layers, $\lambda$ is the wavelength of the short lattice, and $V_{s}$ and $V_{l}$ are the depths of the short and long lattices, respectively, where the wavelength of the long lattice is double the wavelength of the short lattice. 
The interference of the short and long lattice waves produces a potential disparity between neighboring lattice sites.

The dynamics of the model are controlled by the tunneling amplitude $J$, and the on-site interaction $U$, both of which depend primarily on the depth of the optical lattice. Varying the relative depth of the superlattice further controls the offset $\delta$ \cite{Yang2020ObservationGaugeInvariance}.
The lattice tilt $\gamma$ can be implemented via a linear potential acting on the atoms, for example, through gravitational forces or a magnetic-field gradient \cite{Miyake2013RealizingHarperHamiltonian}.

The nearest- and next-nearest-neighbor interactions, with coupling strengths $V$ and $W$ respectively, are realised and controlled by dipole–dipole interactions between highly magnetic atoms \cite{Chomaz2022DipolarPhysicsReview,Lahaye2009PhysicsDipolarBosonic,Griesmaier2005BoseEinsteinCondensationChromium,Lu2011StronglyDipolarBoseEinstein,Aikawa2012BoseEinsteinCondensationErbium}.
The dipole–dipole interaction for parallel dipoles takes the form
\begin{math}
    \hat{V}(\mathbf{r})=\frac{C}{r^{3}}\left(1-3(d\cdot\mathbf{r})^{2}\right),
\end{math}
where $\mathbf{r}$ denotes the relative position vector, $C$ is the effective dipole
moment, and $d$ specifies the polarization direction.
To suppress unwanted long-range matter–matter interactions, we employ a magic-angle
configuration that cancels dipolar couplings along the $x$ and $y$ directions of the
matter sites, realized by applying a magnetic-field gradient to rotate the dipoles to
the magic angle.  
The polarization directions that cancel the dipolar interaction along the $x$ and $y$
directions are $d=\frac{1}{\sqrt{3}}\left(\pm 1, \pm 1, \pm 1\right)$.
Choosing the polarization direction along the z-axis at the gauge sites prevents magic-angle cancellation for interlayer coupling and thus maintains the desired gauge–matter interaction strength.
The condition $V=2W$ determines that the interlayer separation should be $h\approx\frac{a}{2}\sqrt{2^{1/3}-1}\simeq 0.25\,a$.
Due to the dipolar nature of the interaction and the fact that in three dimensions only two directions can be tuned to cancel the coupling, a residual matter--matter interaction remains along other directions, with the most dominant contribution being the nearest neighbor along the diagonal.
Evaluating the ratio between the diagonal matter--matter interaction $V_{m,m}(r_{\text{diag}})$ and the matter--gauge interaction $V_{m,g}$ yields $V_{m,m}(r_{\text{diag}})/V_{m,g} \approx 1/(2^4)\approx0.06$.
While this is small, it is still expected to have a noticeable effect on the resulting dynamics of the system, requiring a more thorough rederivation of the mapping explicitly including longer range interactions for a more realistic setup.
Nevertheless, the slightly simplified model we consider in the present work demonstrates that we can indeed include higher-dimensional gauge field representations in a cold-atom gauge theory simulator.

\begin{figure*}[t]
    \centering
    \includegraphics{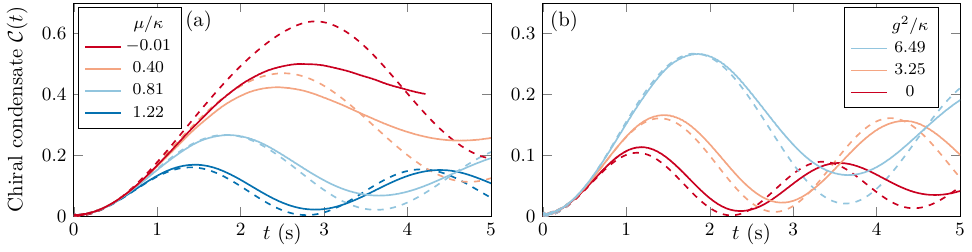}
    \caption{
    Numerical simulations of quenches from the vacuum state in our proposed quantum simulator of the $(2+1)$D spin-1 QLM, showing the evolution of the chiral condensate in both the Bose--Hubbard simulator (solid curves) and the target QLM (dashed curves).
    All simulations are performed on an infinite-length cylinder with a width of $N_y = 2$ matter sites, corresponding to four bosonic sites in the BHM.
    (a)~Quench dynamics for different values of the staggering potential $\delta$, which affects the mass~\(\mu\) in the target model. 
    The on-site interaction strength is fixed to~\(U=160\,\text{Hz}\), while $\delta$ is set to \(79.64\), \(79.59\), \(79.54\), and \(79.49\,\text{Hz}\) (top to bottom).
    (b) ~Quench dynamics for different values of the electric field energy~\(g^2\), with \(\mu\) fixed to approximately \(0.81\kappa\). 
    This is achieved by simultaneously varying the BHM interaction strength~\(U\) to \(160\), \(159.95\), and \(159.9\,\text{Hz}\) (top to bottom) and the corresponding staggering potentials~\(\delta\) to \(79.54\), \(79.465\), and \(79.39\,\text{Hz}\), respectively.
    }
    \label{fig:vac-quench}
\end{figure*}

The ratio $\alpha$ of the on-site interaction strengths can be tuned by exploiting the
different Wannier functions associated with two Bloch bands.
By choosing $V_l$ close to the band gap between the $s$ and $p$ bands of the short lattice, atoms on the matter sites occupy the $s$ band, while atoms on the gauge sites populate the $p$ band.
In the experiment, the ratio $\alpha$ is approximately $\alpha \approx 1.3$.
The potential on the forbidden sites can be created by introducing an additional energy penalty using tightly focused addressing beams.

To prepare the vacuum state as an initial state, we propose, as in Ref.~\cite{Osborne2025Scale$2+1$DGaugeTheory}, to drive the system from a superfluid to a Mott-insulating phase.
This can be achieved by first loading the atoms into the superlattices, cooling the system, and subsequently increasing the lattice depth~\cite{Yang2017SpindependentOpticalSuperlattice}.
The Mott-insulating regime is reached by tuning the system into a parameter regime in which the on-site interaction $U$ exceeds the hopping amplitude $J$, creating a large energy imbalance between neighboring lattice sites and favouring occupation of the deeper gauge sites.
Excess atoms on the matter sites can be removed using an atom-removal operation, resulting in a well-defined initial state \cite{Yang2017SpindependentOpticalSuperlattice}.

\section{Numerical results}
\label{sec:numerical_results}

We benchmark the effectiveness of our proposed quantum simulator for the spin-1 \(\UN(1)\) QLM in two spatial dimensions by computing the time evolution of the BHM~\eqref{eq:H_BHM} and comparing it with that of the target QLM~\eqref{eq:H_QLM}. 
The time evolution is numerically simulated using infinite matrix product state (iMPS) methods \cite{Schollwoeck2005DensityMatrixRenormalizationGroup,Paeckel2019TimeevolutionMethodsMatrixproduct,mptoolkit}, specifically the time-dependent variational principle (TDVP) algorithm \cite{Haegeman2011TimeDependentVariationalPrinciple,Lubich2015TimeIntegrationTensor,Haegeman2016UnifyingTimeEvolution} with single-site updates and adaptive environment expansion \cite{McCulloch2024CommentControlledBond}. 
As is standard with studying systems in two spatial dimension using iMPS techniques, we map the system to to an infinitely long cylinder \(N_x \rightarrow \infty\) with a finite circumference \(N_y\).
To compare the two models throughout the quench dynamics, we use the chiral condensate $\mathcal{C}(t)$, which is defined in the QLM and in the BHM, respectively, as
\begin{subequations}
\begin{align}
    \mathcal{C}_{\qlm}(t)&=
    \lim_{N_{x} \to \infty}\frac{1}{N_{x}N_{y}}
    \sum\limits_{\vb{r}}
    \mel**{\psi(t)}{\Op{\phi}[\vb{r}]* 
    \Op{\phi}_{\vb{r}}}{\psi(t)},
    \\ 
    \mathcal{C}_{\bhm}(t)&=
    \lim_{L_{x} \to \infty}\frac{1}{L_{x}L_{y}}
    \sum\limits_{\vb{j}\in \text{matter}}
    \mel**{\psi(t)}{\Op{n}_{\vb{j}}}{\psi(t)}.
\end{align}
\end{subequations}
Figure~\ref{fig:vac-quench} shows the dynamics of a system initialized in the vacuum state and quenched to different Hamiltonian parameters on a cylinder with a width of $N_y = 2$ matter sites. 
We use the BHM parameters \(J = 1\,\mathrm{Hz}\), \(V = 160\,\mathrm{Hz}\), \(W = 80\,\mathrm{Hz}\), \(\gamma_x = 7\,\mathrm{Hz}\), \(\gamma_y = 5\,\mathrm{Hz}\), \(\alpha = 1.3\), with different values of \(U\) and \(\delta\) cited in the caption of Fig.~\ref{fig:vac-quench}, in order to tune the effective QLM parameters.
In Fig.~\ref{fig:vac-quench}(a), we compare the resulting dynamics in the QLM and the BHM for different values of the ratio $\mu/\kappa$. 
For sufficiently large $\mu/\kappa$, the two models exhibit excellent agreement within the numerically accessible time window.
However, as $\mu/\kappa$ decreases, increasing deviations emerge between the target gauge theory and the dynamics realized in the cold-atom quantum simulator.
This can be understood as follows: Since the effective gauge-theory description is derived perturbatively in $J \propto \kappa$, it is accurate for large $\mu/\kappa$, whereas decreasing $\mu/\kappa$ enhances neglected higher-order processes and leads to growing deviations.
We emphasize that, despite these deviations, the dynamics in the cold-atom quantum simulator remains gauge-invariant, as the tunneling strength $J$ is chosen to be sufficiently small and well within the perturbative regime. 
This ensures that gauge violations remain controlled throughout the evolution, albeit at the cost of slower dynamics.
Similarly, in Fig.~\ref{fig:vac-quench}(b), we show that the simulator is capable of simulating different strengths of the gauge coupling \(g^2/\kappa\), while keeping the mass fixed to \(\mu/\kappa \approx 0.81\).

We can measure local Gauss’s-law violations in the cold-atom quantum simulator using the quantity~\cite{Halimeh2020Fateoflatticegauge}
\begin{align}\label{eq:eta}
    \eta(t)=\lim_{L_{x}\to \infty }\sqrt{\frac{1}{L_{x}L_{y}}
     \sum\limits_{\vb{j}\in \text{matter}}
    \mel**{\psi(t)}{\Op{G}[\vb{j}][2]}{\psi(t)}} 
    .
\end{align}
This quantity is zero for a state in the physical gauge sector, and if it is nonzero, we can use its value to quantify the degree of gauge violation.
In our simulations, we find that this value is well-bounded, staying below \(0.1\%\) for all accessible evolution times.

While the simulations are performed using a system size of \(N_y = 2\), we can also observe similar dynamics for larger system sizes (\(N_y = 4\)) at early times.
However, since the timescales required to obtain coherent dynamics on the Bose--Hubbard simulator are quite large, it is prohibitively expensive to perform such simulations using matrix product state numerics for larger system sizes due to the rapid growth of entanglement.
We also note that we have calculated similar quenches starting from a charge-proliferated state in Appendix~\ref{app:cp}. However, as first-order effects are much stronger there, we do not obtain as close agreement as for the vacuum state, but nevertheless the correct qualitative trend is observed.
It is expected that decreasing the bosonic hopping~\(J\) would reduce such unwanted first-order effects, but at the cost of increasing the timescale of the effective dynamics. The difficulty of classically simulating this dynamics further motivates the importance of our quantum simulation proposal in this work.

\section{Discussion and Outlook}
\label{sec_discussion_outlook}

We have presented a mapping of spin-$S$ $\UN(1)$ quantum link models in $(2+1)$ dimensions onto a bosonic ultracold-atom quantum simulator implemented in an optical superlattice.  
Focusing on current experimental capabilities, we proposed an extended single-species Bose–Hubbard platform based on a bilayer superlattice architecture that spatially separates matter and gauge sites, enabling tunable tunneling and interactions while suppressing unwanted long-range couplings via magic-angle control of dipole-dipole interactions, together with a realistic protocol for state preparation.
Using perturbation theory, we derived an effective Hamiltonian for the BHM describing the QLM and computed relations between the parameters of the quantum simulator and those of the target model. 
Employing matrix product state numerical time evolution simulation, we benchmarked the quench dynamics of local observables, demonstrating good quantitative agreement between the simulator and the quantum link model over all accessible evolution times. 
In all cases, gauge violations were strongly suppressed and well-controlled, a direct consequence of a naturally arising linear gauge-protection term in our mapping. 

This work paves the way for the exploration of larger-spin representations of the quantum link model of QED on modern quantum-simulator platforms.
This is crucial for studying the richer phenomenology beyond the simplest case of the spin-\(1/2\) representation and connecting to the lattice-QED limit.
For example, a recent study~\cite{cao2026stringbreakingglueballdynamics} showed that the spin-\(1\) QLM in two spatial dimensions features mechanisms of string breaking not present in the spin-\(1/2\) formulation, including glueball formation and the realization of strings beyond minimal length.
This scheme for simulating higher-dimensional gauge fields could also be applied in principle to a \((3+1)\)-dimensional extension of this cold atom setup~\cite{orlando2026scalablecoldatomquantumsimulator}, which we leave open for future work. It is important to mention here that even though we employed hard-core bosonic matter, our work can be extended to fermionic matter by, e.g., utilizing alkaline-earth-like atoms in state-dependent optical lattices \cite{Surace2023AbInitioDerivation} or employing Wilson fermions \cite{Zache2018QuantumsimulationoflatticegaugetheoriesusingWilsonfermions}.

\footnotesize
\begin{acknowledgments}
    J.J.O.~and J.C.H.~acknowledge funding by the Max Planck Society, the Deutsche Forschungsgemeinschaft (DFG, German Research Foundation) under Germany’s Excellence Strategy – EXC-2111 – 390814868, and the European Research Council (ERC) under the European Union’s Horizon Europe research and innovation program (Grant Agreement No.~101165667)—ERC Starting Grant QuSiGauge. This work is part of the Quantum Computing for High-Energy Physics (QC4HEP) working group. Numerical simulations were performed on The University of Queensland's School of Mathematics and Physics Core Computing Facility \texttt{getafix}. B.Y.~acknowledges support from the NSFC (Grant No.~12274199), the Shenzhen Science and Technology Program (Grant No.~KQTD20240729102026004), the Guangdong Major Project of Basic and Applied Basic Research (Grant No.~2023B0303000011), and the Guangdong Provincial Quantum Science Strategic Initiative (Grant No.~GDZX2304006, Grant No.~GDZX2405006).
    P.M.~and S.M.~acknowledge the financial support from the European Union’s Horizon 2020 Research and Innovation Programme Grant Agreement No.~101113690 (PASQuanS2) and under the Marie Skłodowska-Curie Grant Agreement No.~101034267 (ENGAGE).
    P.H.~acknowledges support by the European Union under Horizon Europe Programme - Grant Agreement 101080086 - NeQST, CARITRO through project SQuaSH, the Swiss State Secretariat for Education, Research and lnnovation (SERI) under contract number UeMO19-5.1, Provincia Autonoma di Trento, and Q@TN, the joint laboratory between University of Trento, FBK—Fondazione Bruno Kessler, INFN—National Institute for Nuclear Physics, and CNR—National Research Council. \\
    Views and opinions expressed are, however, those of the author(s) only and do not necessarily reflect those of the European Union or the European Research Council Executive Agency. Neither the European Union nor the granting authority can be held responsible for them. 
\end{acknowledgments}
\normalsize

\appendix
\section{Particle hole transformation}
\label{sec:particle_hole_transformation}

When discretizing Dirac fermions on the lattice, the procedure introduces additional unphysical modes in the spectrum, known as doublers.
Various approaches have been developed to address this issue, such as Wilson fermions \cite{Wilson1977QuarksStringsLattice,Zache2018QuantumSimulationLattice}, which remove doublers in arbitrary spatial dimensions at the cost of breaking chiral symmetry.
Here, we employ staggered fermions \cite{Kogut1975HamiltonianFormulationWilsons,Susskind1977LatticeFermions}, where a single spinor field is distributed over two lattice sites. 
This approach partially resolves the doubling problem (with residual doublers in $D>1$) while offering a lower computational cost.
To remove the staggering factors and obtain the final form of our target Hamiltonian in Eq.~\eqref{eq:H_QLM}, we apply a particle–hole transformation 
\cite{Hauke2013QuantumSimulationLattice}:
\begin{align}
    \Op{\phi}_{\vb{r}}
    &\rightarrow 
    \frac{1+(-1)^{r_{x}+r_{y}}}{2} \Op{\phi}[\vb{r}]
    +\frac{1-(-1)^{r_{x}+r_{y}}}{2} \Op{\phi}[\vb{r}]*
    \ ,
    \nonumber\\
    \Op{s}[\vb{r},\vb{e}_{\nu}][+]
    &\rightarrow 
    c_{\vb{r},\vb{e}_{r}}\left[
    \frac{1+(-1)^{r_x+r_y}}{2} \Op{s}[\vb{r},\vb{e}_{\nu}][-]+
    \frac{1-(-1)^{r_x + r_y}}{2} \Op{s}[\vb{r},\vb{e}_{\nu}][+]
    \right]
    \ ,
\end{align}
where $c_{r,e_{r}}$ is a staggering factor \cite{Susskind1977LatticeFermions}. 
This particle–hole transformation removes the staggering factors and enables a consistent mapping of local configurations to bosonic configurations upon adopting a hard-core bosonic representation of the matter fields. 
The resulting QLM Hamiltonian is given in Eq.~\eqref{eq:H_QLM}.

\section{Perturbation theory}
\label{sec:perturbation_theory}

\begin{figure}[!t] 
  \centering
  \begin{minipage}[t]{0.47\columnwidth}
    \centering
    \resizebox{\linewidth}{!}{\begin{tikzpicture}[scale=1]

\def\rows{5}
\def\cols{5}

\foreach \i in {0,...,4} {
  \foreach \j in {0,...,4} {
    \coordinate (P) at (\j, \i);
    
    \draw[black,line width=2] (\j, \i - 0.6) -- ++(0,1);
    \draw[black,line width=2] (\j - 0.6, \i) -- ++(1,0);
    
    \draw[-{Latex[length=3mm]}, line width=2pt] (\j-0.7,\i) -- (\j-0.2,\i); 
  }
}

\foreach \i in {0,...,4} {
  \foreach \j in {0,...,4} {
    \draw[fill=white] (\j,\i) circle (3pt);
  }
}

\end{tikzpicture}}
    \medskip
    \small (a) 
  \end{minipage}
  \hspace{0.001\columnwidth}
  \vrule width 0.5pt
  \hspace{0.02\columnwidth}
  \begin{minipage}[t]{0.47\columnwidth}
    \centering
    \resizebox{\linewidth}{!}{\begin{tikzpicture}[scale=1]

\foreach \i in {0,...,4} {
  \foreach \j in {0,...,4} {

    \draw[black,line width=2] (\j, \i - 0.6) -- ++(0,1);
    \draw[black,line width=2] (\j - 0.6, \i) -- ++(1,0);

    \ifnum\j=0
    \else
      \ifnum\j=2
      \else
        \ifnum\j=4
        \else
          \draw[-{Latex[length=3mm]}, line width=2pt] (\j-0.7,\i) -- (\j-0.2,\i);
        \fi
      \fi
    \fi

  }
}

\foreach \i in {0,...,4} {
  \foreach \j in {0,...,4} {
    \draw[fill=black] (\j,\i) circle (3pt);
  }
}

\end{tikzpicture}}
    \medskip
        \small (b) 
  \end{minipage}
  \caption{(a)~Vacuum state in the QLM with an energy density of $2g^2$ per four-site unit cell. 
    (b)~Charge-proliferated state with an energy density of $4\mu + g^2$ per four-site unit cell.
    }
  \label{fig:charge_prof_states}
\end{figure}
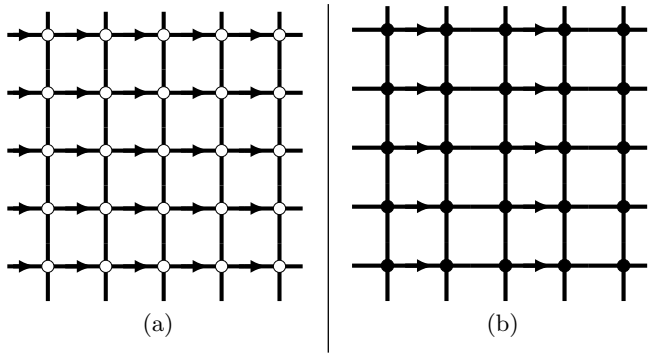

To establish the relationship between the parameters of the target Hamiltonian $\Op{H}[\qlm]$, namely the tunneling constant $\kappa$, the mass parameter $\mu$, and the gauge coupling $g$, and those of the extended BHM, we derive an effective Hamiltonian using second-order perturbation theory via the Schrieffer–Wolff transformation \cite{Schrieffer1966RelationAndersonKondo,Bravyi2011SchriefferWolffTransformation}.
To stabilize gauge configurations via a linear protection term, we introduce a potential gradient through the tilt vector $\gamma$. 
To additionally suppress unwanted resonances in the spectrum, the gradient is chosen to differ slightly between the $x$ and $y$ directions, yielding two distinct functions for each parameter, $g_{x,y}$, $\mu_{x,y}$, and $\kappa_{x,y}$.
These functions labeled with $x$ and $y$ are obtained from second-order processes along the $x$- and $y$-axes, respectively.
Specifically, we decompose the BHM Hamiltonian~\eqref{eq:H_BHM} into a diagonal contribution and a purely off-diagonal term in the boson occupation-number basis,
\begin{subequations}
\begin{align}
    \Op{H}[\mathrm{diag}]=&\sum_{\vb{j}} 
    \frac{U_{\vb{j}}}{2} 
    \Op{n}_{\vb{j}}\left(\Op{n}_{\vb{j}}-1\right)
    +\sum_{\vb{j}} 
    \left(\boldsymbol{\gamma}\cdot\vb{j}-\delta_{\vb{j}}\right) \Op{n}_{\vb{j}}
    \nonumber\\
    &+ V \sum_{\vb{j},\nu}
    \Op{n}_{\vb{j}} \Op{n}_{\vb{j}+\vb{e}_{\nu}} 
    +W \sum_{\vb{j},\{\nu,\nu'\}\in\Omega}
    \Op{n}_{2\vb{j}+\vb{e}_{\nu}} \Op{n}_{2\vb{j}+\vb{e}_{\nu'}},
    \\
    \Op{H}[\mathrm{off-diag}]=&
    -J \sum_{\vb{j},\nu}
    \left(\Op{b}^{\dagger}_{\vb{j}} \Op{b}_{\vb{j}+\vb{e}_{\nu}}+\mathrm{H.c.}\right).
\end{align}
\end{subequations}
Subsequently, we perform a unitary transformation on the BHM Hamiltonian, with $\Op{U}$ defined as $\Op{U}\equiv e^{-i\Op{S}}$,
\begin{equation}
    \Op{H}[\bhm]'=e^{i\Op{S}}\,\Op{H}[\bhm]\, e^{-i\Op{S}}
    \ ,
\end{equation}
such that the resulting effective Hamiltonian $\Op{H}[\bhm]'$ is diagonal up to second order in $\Op{H}[\bhm,\mathrm{off-diag}]$. 
To achieve this, we define the generator $\Op{S}$ of the unitary transformation such that
\begin{equation}
    [\Op{S},\Op{H}[\mathrm{diag}]]
    =i\Op{H}[\mathrm{off-diag}]
    \ .
\end{equation}
Thus, the effective Hamiltonian up to the second order in perturbation theory reads
\begin{align}
    \Op{H}[ij]'=\delta_{ij} E_i+&\frac{1}{2}\sum_k
    \left(\frac{1}{E_i-E_k}+\frac{1}{E_j -E_k}\right) 
    \nonumber\\
    &\times \bra{i}\Op{H}[\mathrm{off-diag}] \ket{k} 
    \bra{k}\Op{H}[\mathrm{off-diag}] \ket{j}
    \ ,
    \label{eq:H_eff}
\end{align}
where $E_{i}=\langle i | \Op{H}[\bhm,\mathrm{diag}]| i \rangle$ are the eigenvalues of the diagonal part of the Hamiltonian.
For the perturbative calculation, we consider all possible states within a translation-invariant unit cell.
As the unit cell, we define a single plaquette of the QLM with PBC.
This choice guarantees that bosonic configurations in the BHM have the same energy if and only if the corresponding QLM configurations have the same energy.
In the effective description~\eqref{eq:H_eff}, bosonic configurations associated with different gauge-invariant states of the unit cell are connected through effective first-order processes that arise at order $\mathcal{O}(J^2)$.
To establish the mapping between the Hamiltonian coefficients of the two theories, we compare the QLM energy density per unit cell with the corresponding matrix elements of the extended BHM obtained from Eq.~\eqref{eq:H_eff}.
As the state with $E_{\qlm}=0$ does not correspond to a zero-energy state in the BHM, we consider relative energy differences instead.

Expressions for $g^2$ and $\mu$ are obtained by analysing the matrix elements of two different states shown in Fig.~\ref{fig:charge_prof_states}, which yield a system of equations for the coefficients.
From the state in Fig.~\ref{fig:charge_prof_states}(a), we compute the corresponding diagonal matrix element using Eq.~\eqref{eq:H_eff}, subtract the vacuum energy, and thereby extract $g^2$.
For the state in Fig.~\ref{fig:charge_prof_states}(b), which has an energy density of $4\mu + g^2$ in the QLM, we compute the corresponding BHM matrix element to second order. 
Substituting the previously determined value of $g^2$ and subtracting the BHM vacuum energy then yields the parameter $\mu$.

\begin{figure}[t]
    \centering
    \scalebox{0.43}{\tikzset{every picture/.style={line width=0.75pt}} 

\begin{tikzpicture}[x=0.75pt,y=0.75pt,yscale=-1,xscale=1]
\def\linethick{5} 
\def\circrad{12} 
\def\circradb{10} 

\draw[line width=\linethick pt](152,120) -- (258,120) ;
\draw[line width=\linethick pt](282,120) -- (330,120) ;
\draw[loosely dashed,line width=\linethick pt,line cap=round]  (80,120) -- (127,120) ;
 
\draw[line width=\linethick pt](152,230) -- (258,230) ;
\draw[line width=\linethick pt](282,230) -- (330,230) ;
\draw[loosely dashed,line width=\linethick pt,line cap=round](80,230) -- (127,230) ;

\draw[line width=\linethick pt](470,120) -- (580,120) ;
\draw[line width=\linethick pt](602,120) -- (650,120) ;
\draw[loosely dashed,line width=\linethick pt,line cap=round](400,120) -- (447,120) ;

\draw[
  line width=\linethick pt,
  postaction={decorate},
  decoration={
    markings,
    mark=at position 0.45 with {\arrow[scale=1.6]{Latex[length=5mm, width=4mm]}}
  }
] (470,230) -- (650,230);

\draw[loosely dashed,line width=\linethick pt,line cap=round]  (400,230) -- (454,230) ;

\draw[line width=\linethick pt](140,50) -- (140,108) ;
\draw[line width=\linethick pt](140,130) -- (140,220) ;
\draw[loosely dashed,line width=\linethick pt,line cap=round](140,244) -- (140,290) ;

\draw[line width=\linethick pt](270,50) -- (270,108) ;
\draw[line width=\linethick pt](270,132) -- (270,220) ;
\draw[loosely dashed,line width=\linethick pt,line cap=round] (270,244) -- (270,290) ;

\draw[line width=\linethick pt](460,50) -- (460,108) ;
\draw[line width=\linethick pt](460,132) -- (460,230) ;
\draw[loosely dashed,line width=\linethick pt,line cap=round](460,244) -- (460,284) ;

\draw[line width=\linethick pt](590,50) -- (590,108) ;
\draw[line width=\linethick pt](590,130) -- (590,230) ;
\draw[loosely dashed,line width=\linethick pt,line cap=round](590,244) -- (590,284) ;

\draw[line width=0.5 pt] (365,10) -- (365,320);

\draw [black, line width=1] (270,230) circle (\circrad);
\draw [black, line width=1] (140,120) circle (\circrad);
\draw [black, line width=1] (270,120) circle (\circrad);
\draw [black, line width=1] (140,230) circle (\circrad);

\draw [fill=black, fill opacity=1] (460,230) circle (\circradb);
\draw [fill=black, fill opacity=1] (590,230) circle (\circradb);
\draw [black, line width=1] (590,120) circle (\circrad);
\draw [black, line width=1] (460,120) circle (\circrad);

\node[anchor=mid] at (205,20) {\Huge$E_{\mathrm{QLM}} = 0$};
\node[anchor=mid] at (530,20) {\Huge$E_{\mathrm{QLM}} = 2\mu+\frac{g^2}{2}$};

\node at (205,310) {\Huge (a)};
\node at (530,310) {\Huge (b)};

\end{tikzpicture}}
    \caption{(a)~Vacuum state of the QLM. 
    (b)~Physical state with an energy density of $2\mu + g^2/2$. 
    By calculating the matrix element between these two states, which are connected via a second-order process, an expression for $\kappa_x$ is obtained.
    }
    \label{fig:states_kappa}
\end{figure}
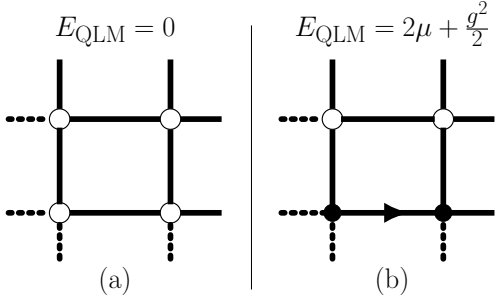

Finally, to obtain $\kappa_{x}$, we evaluate the matrix elements of the states depicted in Fig.~\ref{fig:states_kappa}. 
Analogous states are then used to extract the coefficient $\kappa_{y}$.
The resulting expression can be simplified using the resonance conditions in Eqs~\eqref{eq:resonance_A} and \eqref{eq:resonance_B}, which reduce to
\begin{subequations}
\begin{align}
    U &\approx 2W\, , \\
    6V &\approx 13W - \delta\, .
\end{align}
\end{subequations}
Together with the condition \(V = 2W\), which prevents unwanted long-range coupling terms from appearing in the mapped model, this allows us to simplify the expressions and reduce the number of variables.

The matrix elements required to derive the relations between the Hamiltonian coefficients are obtained by applying Eq.~\eqref{eq:H_eff} to the corresponding states in the bosonic realization and are presented below:
\begin{widetext}
\begin{subequations}
\begin{align}\label{eq:matrix_element_vac}
    \bra{0}\Op{H}\ket{0}&=8 U + 96 W - 16 \delta 
    +\sum_{\gamma \in \{\gamma_x,\gamma_y\}}
    \left(\frac{8 J^{2}}{U - 7 V + 12 W - \delta + \gamma} + (\gamma\to-\gamma)\right),\\\label{eq:matrix_element_g2}
    \bra{2g^2}\Op{H}\ket{2g^2}&=
    16 U + 80 W - 16 \delta
    +\sum_{\gamma \in \{\gamma_x,\gamma_y\}}
    \left(\frac{8 J^{2}}{U - 7 V + 12 W - \delta + \gamma} + (\gamma\to-\gamma)\right),\\\nonumber
    \bra{4\mu+g^2}\Op{H}\ket{4\mu+g^2}&=
    6 U + 24 V + 48 W - 12 \delta
    +\left(\frac{16 J^{2}}{(1 - \alpha)U - 3 V + 8 W - \delta + \gamma_{y}} + (\gamma_{y}\to-\gamma_{y})\right)
    \nonumber\\
    &+ \left(\frac{8 J^{2}}{(1- \alpha) U - 3 V + 8 W - \delta + \gamma_{x}} +  (\gamma_{x}\to-\gamma_{x})\right)
    + \left(\frac{2 J^{2}}{5 V - 12 W + \delta + \gamma_{x}} + (\gamma_{x}\to-\gamma_{x})\right)
    \nonumber\\
    &+ \left(\frac{12 J^{2}}{- 2 U + 5 V - 8 W + \delta + \gamma_{y}} +(\gamma_{y}\to-\gamma_{y})\right)
    +\left( \frac{6 J^{2}}{- 2 U + 5 V - 8 W + \delta + \gamma_{x}} +(\gamma_{x}\to-\gamma_{x})\right),
    \nonumber\label{eq:matrix_element_4mu_g2} \\
    \bra{2\mu+g^2/2}\Op{H}\ket{0}&=
    \left( \frac{J^{2}}{5 V - 12 W + \delta + \gamma_{x,y}}
    + (\gamma_{x,y} \rightarrow -\gamma_{x,y}) \right) 
    + \left(
    \frac{J^{2}}{U - 7 V + 12 W - \delta + \gamma_{x,y}} 
    + (\gamma_{x,y} \rightarrow -\gamma_{x,y}) \right),
\end{align}
\end{subequations}
\end{widetext}
where, as mentioned in the main text, we use the notation \((\gamma_{i} \rightarrow -\gamma_{i})\) to denote repeating the previous expression but with \(\gamma_{i}\) replaced by \(-\gamma_{i}\).
The matrix element $\bra{0}\Op{H}\ket{0}$ gives the diagonal contribution of the state up to second order and corresponds to the vacuum state in the QLM.
In contrast, the matrix elements $\bra{2g^2}\Op{H}\ket{2g^2}$ and $\bra{4\mu+g^2}\Op{H}\ket{4\mu+g^2}$ correspond to the states shown in Fig.~\ref{fig:charge_prof_states}(a) and (b), respectively.
The states used to evaluate the off-diagonal matrix element $\bra{2\mu+g^2/2}\Op{H}\ket{0}$ are shown in Fig.~\ref{fig:states_kappa}(a) and (b), from which the coefficient $\kappa_x$ is subsequently derived. 
An analogous process along the $y$ axis yields the Hamiltonian coefficient $\kappa_{y}$.

\section{Charge proliferated state}
\label{app:cp}

\begin{figure}[t]
    \centering
    \includegraphics{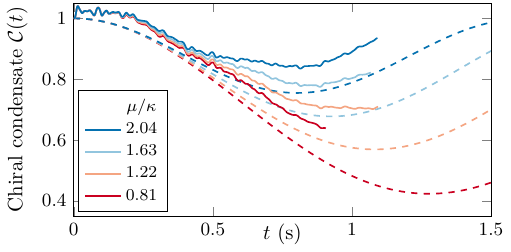}
    \caption{
    Numerical simulation of quenches from the charge-proliferated product state in our proposed $(2+1)$D spin-1 QLM quantum simulator, showing the evolution of the chiral condensate in the Bose--Hubbard simulator (solid curves) and the target QLM (dashed curves).
    We keep the on-site interaction strength~\(U\) fixed to \(160\,\text{Hz}\), setting the staggering potential~\(\delta\) to be \(79.39\), \(79.44\), \(79.49\), and \(79.54\,\text{Hz}\) (top to bottom).}
    \label{fig:cp-quench}
\end{figure}

In this section, we benchmark the mapping by performing quenches starting from a charge-proliferated state, rather than a vacuum state as considered in the main text.
The results are presented in Fig.~\ref{fig:cp-quench}.
Although the simulator presents roughly similar qualitative behavior to the desired QLM dynamics, the deviation is much stronger than in the case of the quenches of the vacuum state.
Indeed, the gauge violation \(\eta(t)\)~\eqref{eq:eta} is much larger for these quenches, and rises to approximately \(1.2\%\).
In this case, we can see that even at very early times, first-order effects are much stronger here, as bosonic hopping processes from gauge sites to occupied matter sites have an amplified energy.
While it would be expected that such unwanted first-order dynamics would become less relevant as we decrease the bosonic hopping \(J\), this would also result in the effective dynamics taking place over a much longer timescale, beyond the reach of classical numerical simulation.

\section{Including forbidden sites}
\label{sec:forbidden_sites}

In this section, we estimate the correction to the effective mass arising from retaining the forbidden sites in the lattice.
In this case, a chemical potential of strength $\eta \gg \delta$ is used to penalize occupation of the forbidden sites.
As a result, single bosons are no longer restricted to hopping exclusively between matter and gauge sites, but may also occupy the forbidden sites with a suppressed probability.
The correction is computed using the same procedure described previously, now including first-order processes in which bosons hop to forbidden sites.
For typical parameter values $U \approx V \approx 2\delta$, $W \approx \delta$, and $\gamma_{x,y} \approx 0$, the resulting mass correction takes the form
\begin{equation}
    \mu_{\text{corr.}} =-2J^2
    \left(
    \frac{1}{7\delta+\eta}
    + \frac{3}{-\delta+\eta}
    + \frac{6}{-9\delta+\eta}
    \right).
\end{equation}
We now consider the case in which the forbidden sites are retained in the optical lattice. 
Their slowly decaying influence can then significantly affect the system dynamics, leading to a renormalization of the effective masses in the QLM.
Although we typically operate in the regime $\delta \ll \eta$, this calculation shows that these processes are not entirely negligible.
Moreover, from an experimental perspective, achieving such large chemical potential differences on individual sites presents a considerable challenge.

\bibliography{biblio}

\end{document}